\documentclass[a4paper,11pt]{article}
\pdfoutput=1 

\usepackage{jcappub} 

\usepackage[english]{babel}
\pdfoutput=1

\usepackage{amsmath,amssymb,mathtools,tabu}
\usepackage{graphicx}
\usepackage{slashed}    
\usepackage{xfrac}                      
\usepackage{url}
\usepackage{color}
\usepackage{multirow}
\usepackage{placeins}
\usepackage[dvipsnames]{xcolor}

\allowdisplaybreaks
\newcommand{\dNeff}{\Delta N_{\rm eff}}
\newcommand{\lcdm}{$\Lambda$CDM }
\newcommand{\lbdm}{$\Lambda$BDM }
\newcommand{\dd}{\delta_{\rm B}}
\newcommand{\dc}{\delta_{\rm c}}
\newcommand{\td}{\theta_{\rm B}}
\newcommand{\fd}{f_{\rm BDM}}
\newcommand{\as}{a_*}
\newcommand{\ts}{\tau_*}
\newcommand{\zs}{z_*}
\newcommand{\ks}{k_*}
\newcommand{\ds}{\Delta}
\newcommand{\dz}{\Delta z/\zs}
\newcommand{\xs}{x_*}
\newcommand{\kt}{k\tau}

\newcommand{\teq}{\tau_{\rm eq}}
\newcommand{\zeq}{z_{\rm eq}}
\newcommand{\tdm}{T_{\rm DM}}
\newcommand{\tcdm}{T_{\rm C}}
\newcommand{\tbdm}{T_{\rm B}}

\abstract{Dark matter may have been relativistic and collisional until
  relatively late times and  become cold and collisionless after a phase
  transition before the matter-radiation equality of the standard
  $\Lambda$CDM cosmology. We show that such a dark matter has large
  peculiar velocities due to acoustic oscillations before the phase
  transition, and evolves ballistically  after the phase transition in the
  collisionless phase until the initial acoustic velocities are redshifted
  away. We show that this Ballistic Dark Matter (BDM) results in new
  non-trivial interesting features in the cosmological observables. In
  particular, the linear matter power spectrum exhibits acoustic
  oscillations on scales smaller than the Hubble scale at the time of phase
  transition, and for fast transitions the power at the acoustic peaks in
  the matter power	spectrum \emph{exceeds} that in a $\Lambda$CDM
  cosmology. If BDM only forms a part of the total dark matter, an odd
  vs. even acoustic peak asymmetry becomes prominent. We give an
  approximate analytical treatment of the linear perturbations in BDM,
  explaining these features. We also discuss  the possibility to constrain
  BDM using cosmic microwave background and large scale structure data.}

\subheader{\hfill TIFR-TH-18-42}

\begin{document}

\title{\huge{Ballistic Dark Matter oscillates above {\boldmath${\mathsf \Lambda}$}CDM}}

\author{Anirban Das,}
\emailAdd{anirbandas@theory.tifr.res.in }
\author{Basudeb Dasgupta,}
\emailAdd{bdasgupta@theory.tifr.res.in}
\author{and Rishi Khatri}
\emailAdd{khatri@theory.tifr.res.in}
\affiliation{Tata Institute of Fundamental Research,
             Homi Bhabha Road, Mumbai, 400005, India.}
\maketitle

\section{Introduction}
Diverse lines of cosmological evidence suggest that approximately $84\%$ of the total matter in the Universe behaves as if it were made of particles without appreciable interactions, either among themselves or with other particles, and with very small velocity dispersion, i.e., collisionless cold dark matter (CDM)~\cite{Ade:2015xua, Costanzi:2018xql, 2012PhRvD..86j3518P, Tegmark:2006az, Tegmark:2003ud, Clowe:2006eq, Begeman:1991iy}. In most particle physics models of dark matter, the dark component could not have been CDM-like at arbitrarily high redshifts. At early enough times, the dark component in most dark matter models would have been relativistic and collisional, coupled to itself and perhaps also with the visible sector. The dark component would become non-relativistic at some epoch and also (almost) collisionless, as the different interaction rates become slower than the expansion rate of the Universe, leading to chemical freeze-out and kinetic decoupling. These three transitions, viz., becoming non-relativistic, freeze-out of interactions with itself, and freeze-out with the visible sector, may or may not always coincide (See Ref.~\cite{Heeba:2018wtf} for an example). 

In the standard WIMP (weakly interacting massive particle) models, freeze-out of dark matter self-annihilations into visible sector particles, which fixes the dark matter abundance, happens long after the dark component is already non-relativistic. See Ref.~\cite{Steigman:2012nb} for an overview. On the other hand, in pure dark gauge sector models, dark gluons decouple from the visible sector very early and undergo a phase transition to form nonrelativistic dark glueballs, whose self-interactions may or may not be important~\cite{Carlson:1992fn,Okun:1980kw,Faraggi:2000pv,Juknevich:2009ji, Acharya:2017szw}. Similarly, in asymmetric dark matter models with light dark quarks, the dark relativistic plasma condenses into nonrelativistic dark baryons which act as the dark matter~\cite{NUSSINOV198555,BARR1990387,Barr:1991qn,PhysRevLett.68.741,Kaplan:2009ag,Kribs:2009fy,Blennow:2010qp}. We will club these theories, where a self-collisional dark radiation transitions to collisionless dark matter at some late redshift $\zs$, long after decoupling from the visible sector,  under the rubric of ``Ballistic Dark Matter'' (BDM). Here \emph{ballistic} refers to the fact that the dark matter at the beginning of the non-relativistic collisionless phase has large initial velocities inherited from the acoustic oscillations in the relativistic collisional phase.

In the BDM model, {the time and duration of the phase transition would
  affect the background cosmology as well as perturbations. If the dark component of the Universe was relativistic at
the time of big bang nucleosynthesis (BBN), it would contribute to radiation
energy density and expansion rate of the Universe. Extra relativistic
degrees of freedom, parameterized by $\dNeff$ (the extra neutrino degrees
of freedom having same energy density), would modify the primordial
nucleosynthesis if the phase transition takes place after the BBN epoch~\cite{2011PhLB..701..296M}.  The cosmic microwave
background (CMB) data suggest that during the epoch of recombination most
of the matter must have been CDM-like~\cite{Ade:2015xua} and restricts the
phase transition to happen before the recombination epoch. We could
 do slightly better and  require  that the phase transition happens before the  redshift of matter-radiation equality, $\zeq$, so that we do not
 change $\zeq$.  The background
 cosmological parameters, like the relative matter and radiation densities
 and the Hubble parameter, would go to the $\Lambda$CDM values after the
 phase transition. The effect on the
 cosmological observables, the CMB and matter power
 spectrum~\cite{Ade:2015xua,Percival:2006gt}, would therefore appear dominantly through the modification of the dark
 matter perturbations for modes which entered the horizon before the phase
 transition. We will see the  small scale modes, which entered the
 horizon before the $\Lambda$CDM matter-radiation equality, and which are
 well measured by CMB and galaxy surveys, would yield the strongest 
 limits on the parameters of the phase transition.}

\begin{figure}
 \begin{center}
  \includegraphics[width=0.55\textwidth]{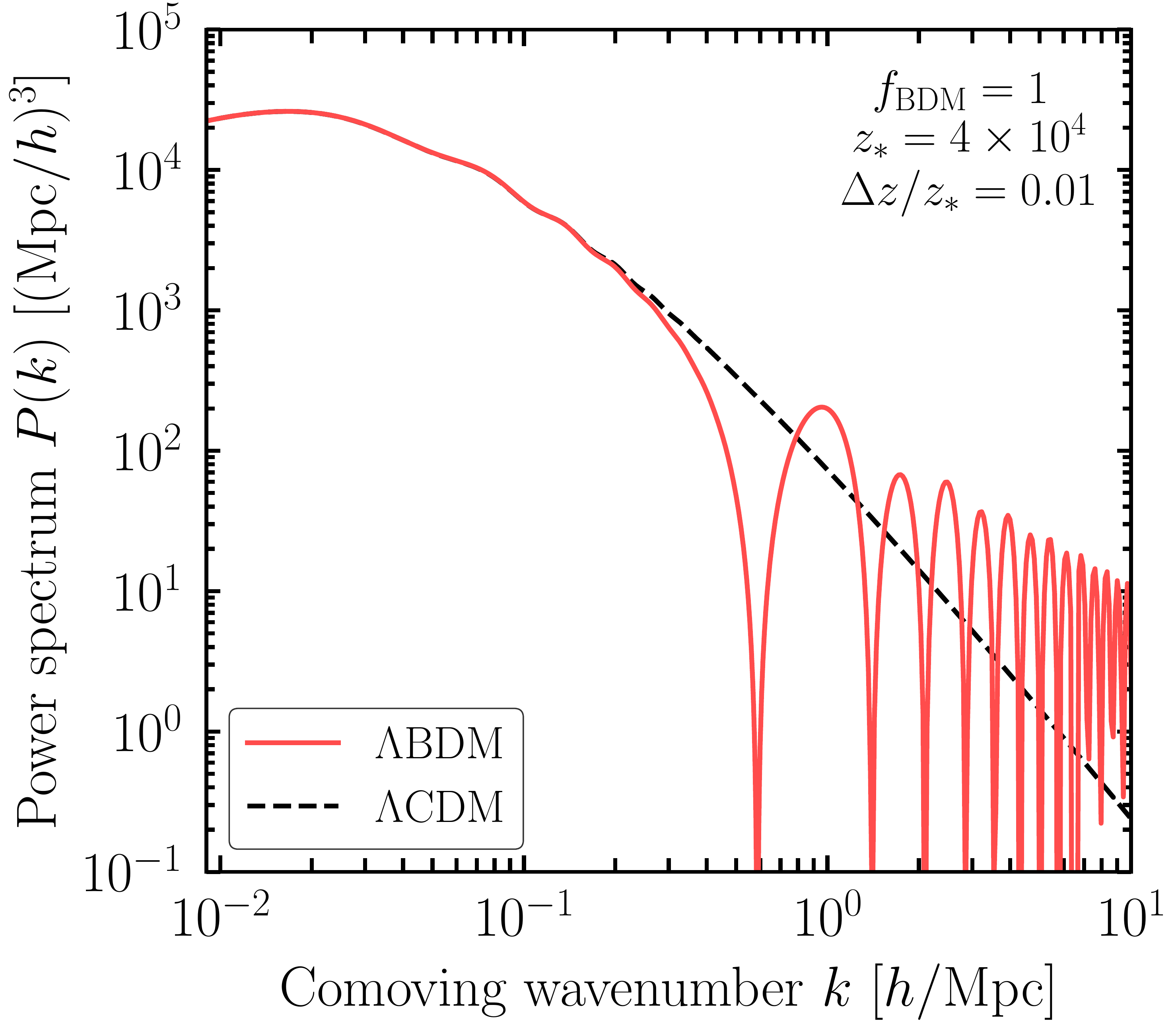}
  \caption{The matter power spectrum (solid red) as predicted in the \lbdm
    cosmology with a fast phase transition at redshift $\zs=4\times 10^4$,
    and BDM comprises all of the dark matter. The \lcdm power spectrum is also
    shown for comparison (dashed black). \label{fig:intro_fig}}
 \end{center}
\end{figure}

The initial evolution of the matter perturbations after the phase
transition is ballistic and results in interesting deviations in the
perturbations compared to the CDM of the standard cosmological model. A
sample of this behavior is shown in Fig.\,\ref{fig:intro_fig}, for a phase
transition at the redshift $\zs=4\times 10^4$ happening over a
redshift-range $\Delta z = 0.01 \zs$ {for BDM forming  all of dark
  matter. We will also consider cosmologies where BDM forms only a fraction
  $f_{\rm BDM}$ of dark matter and the remaining  $1-f_{\rm
    BDM}$ fraction is contributed by the standard CDM. } As one can see,
not only does the matter power spectrum exhibit acoustic oscillations at
small scales, the power at the acoustic peaks exceeds the power in a
standard \lcdm model. The physics of BDM acoustic peaks is similar to
  that of baryon acoustic oscillations (BAOs)~\cite{sz1970,eh1998}.  We will see that, similar
  to the velocity overshoot effect in BAOs which results in phase shift in
  the BAO peaks w.r.t the acoustic peaks in the CMB, the origin of the BDM acoustic peaks
  lies with the velocity perturbations at the time of phase transition and they
  correspond to the extrema of the velocity perturbations.  Both the
  negative minima
   and the positive maxima  of the BDM transfer function
  give rise to the peaks in the power spectrum.  If not all of dark matter
  is BDM, the alternate BDM transfer function extrema would have  either same or opposite sign
  to that of the CDM transfer function. The two DM components would
  alternately add
   constructively and destructively for successive extrema giving prominent asymmetry
  in the heights of odd vs even acoustic peaks in the total matter power
  spectrum. The features we see are similar to the charged massive particle model in Refs.~\cite{Kamada:2016qjo,  Kamada:2017oxi} also studied in Ref.~\cite{Sarkar:2017vls} and have similar origin, namely, the initial
velocity perturbations derived from a previous acoustic phase.  These
features are distinct from those of the other nonstandard models for which
the growth of perturbations has been studied in detail~\cite{Boehm:2000gq,
  Boehm:2001hm, Chen:2002yh, Sigurdson:2003vy, Nusser:2004qu, Das:2006ht,
  Aarssen:2012fx, Wilkinson:2013kia,Wilkinson:2014ksa,Chu:2014lja,
  Buckley:2014hja,  Archidiacono:2015oma, Cyr-Racine:2015ihg,
  Chacko:2016kgg, Dror:2017gjq, Buen-Abad:2018mas}. The
purpose of this paper is to introduce this BDM model, highlight its
predictions, and explain them analytically. {We will also discuss
qualitative constraints on this model from existing CMB and galaxy surveys
but leave a more detailed parameter space study for future work.}


\section{Ballistic Dark Matter: Model \emph{\&} Methods}

We consider an effective macroscopic model for the dark sector, remaining agnostic to its detailed particle physics underpinnings. 
We assume that at early times the dark sector comprises of self-interacting relativistic species, that we call as the dark radiation (DR) phase, and it transitions to  non-interacting non-relativistic particles, i.e.,  dark matter (DM) phase,  at a redshift $\zs$ with the corresponding scale factor denoted by $\as$. We further assume that  the anisotropic stress and all higher order moment perturbations in the Boltzmann hierarchy vanish, leaving only the density and velocity perturbations and allowing the DR phase to be described by its equation of state.  This assumption can be relaxed, at the cost of working with the full stress-energy tensor~\cite{Hu:1998kj}.  In the cosmological context, such an evolution can be encoded in a time-varying equation of state (EoS) parameter for the BDM fluid,
\begin{equation}
	w_{\rm B}(z)=
	\begin{cases}
		\sfrac{1}{3}\quad\,\, z\gg \zs~~({\rm before\,\,transition})\\
		0\quad\quad z\ll \zs~~({\rm after\,\,transition})\,.\\
	\end{cases}
\end{equation}
The subscript ${\rm B}$ will denote quantities associated to the BDM fluid.

Exactly how the EoS transitions between these two limits would depend
  on the details of the particle physics model of BDM. We expect that the
  cosmological observables, like the matter power spectrum, would be
  sensitive to the time or redshift of phase transition and how long the
   transition period lasts. Based on these considerations, we adopt the following 
   simple model for the EoS during the phase transition,
\begin{equation}
	w_{\rm B} = \dfrac{1}{6}\left[1-\tanh\left(\dfrac{a-\as}{\ds}\right)\right]\,,
\label{eq:eos}
\end{equation}
 where $\ds$ parametrizes the extent in scale factor over which the transition takes place. 
Note that as long as $\ds \ll \as$ and $\as \ll 1$, we have
$|\ds/\as| \approx |\dz|$, {where $\Delta z$ is the redshift-duration of
  phase transition.}
We will often refer to redshift instead of scale factor to give a more intuitive sense of when 
the transition takes place and how long it lasts.
 
In general, BDM may form only a fraction $\fd$ of the total dark matter energy density. To parametrize this possibility, we
define
\begin{equation}
	\fd = \dfrac{\Omega_{\rm BDM}}{\Omega_{\rm BDM}+\Omega_{\rm CDM}}\,,
\end{equation}
as the present-day ratio of BDM to total dark matter, with the remaining fraction being CDM-like at all epochs.

We will also be interested in how perturbations grow in the presence of BDM. The equations for the linear perturbations of the BDM fluid in the \emph{conformal Newtonian gauge} can be written as:
\begin{align}
\dot{\dd} &= -(1+w_{\rm B})(\td + 3\dot{\phi}) - 3\dfrac{\dot{a}}{a}\left(c_s^2 - w_{\rm B}\right)\dd\,, \vspace{0.2cm}\label{eq:pert_gen}\\
\dot{\td} &= -\dfrac{\dot{a}}{a}(1-3w_{\rm B})\td - \dfrac{\dot{w}_{\rm B}}{1+w_{\rm B}}\td + \dfrac{c_s^2}{1+w_{\rm B}}k^2\dd + k^2\psi\,.\label{eq:theta}
\end{align}
Here $\dd\equiv\delta\rho_{\rm B}/\rho_{\rm B}$ is the fractional density
perturbation, $\td\equiv ikv_{\rm B}$ is the velocity perturbation
divergence, and $c_s$ is the speed of sound in the BDM fluid. The scalar
metric perturbations are $\phi$ and $\psi$. We follow the sign convention
of Ref.~\cite{Dodelson:2003ft} for $\phi$, which differs by a sign from
Ref.~\cite{Ma:1995ey}. All derivatives are w.r.t. the conformal time
$\tau$. The second term on the RHS of the first equation vanishes because we assume that the EoS does not depend on the energy density,
and hence the speed of sound $c_s^2\equiv \delta P/\delta\rho=w$.  {We  can combine Eqs. (\ref{eq:pert_gen}) and (\ref{eq:theta}) to get a second order equation
  for $\dd$.
\begin{equation}
\begin{array}{l}
\ddot{\dd} +\dfrac{\dot{a}}{a}\left(1-3w_{\rm B}\right)\dot{\dd}+k^2w_{\rm
  B}\dd=F\,,\\\vspace*{-3mm}\\
F\equiv -3\dfrac{\dot{a}}{a}\left(1-3w_{\rm B}\right)\left(1+w_{\rm
    B}\right)\dot{\phi}-3\dot{w}_{\rm B}\dot{\phi}-3\left(1+w_{\rm
    B}\right)\ddot{\phi}
-\left(1+w_{\rm
    B}\right)k^2\psi\,.\label{eq:osc}
\end{array}
\end{equation}
This is a damped forced oscillator equation for $\dd$, with a forcing term
$F$ arising from metric perturbations which would be sourced by the self
gravity of BDM as well as the other components of the Universe. Once the
phase transition starts, the equation of state $w_{\rm B}<1/3$ and  the
damping term ($\propto \dot{\dd}$) becomes non-zero, damping the acoustic
oscillations until the transition to DM phase is complete and the BDM fluid
becomes non-interacting.
}

For the numerical results shown in this paper,  we have implemented the BDM
species, defined by Eqs.\,(\ref{eq:eos} - \ref{eq:theta}), in the public codes {\tt
  CAMB}~\cite{Lewis:1999bs} and {\tt CLASS}~\cite{2011JCAP...07..034B}. We
then computed the transfer functions and the power spectra for this model
using both codes and obtained essentially identical results. When using the
synchronous gauge, we always keep a trace amount of ordinary CDM component
to ensure that the gauge is well-defined. We will assume that the
{stress-energy tensor  components remain continuous across the phase
  transition to connect the DR phase perturbations with the DM phase
  perturbations.} We will also assume that no additional perturbations are
created due to the phase transition.

\section{Cosmological Signatures}

\subsection{Extra Relativistic Degrees of Freedom}

The BDM before the phase transition acts like dark radiation and would contribute to the expansion of the Universe
in the radiation dominated era modifying the expansion rate of the
  Universe. This effect can be quantified by effective relativistic
degrees of freedom (in addition to photons),  defined as
\begin{align}
N_{\rm eff} & = \dfrac{\sum\rho_{\nu_i}}{\rho_{\nu}^{\rm FD}}+\dfrac{\rho_{\rm DR}}{\rho_{\nu}^{\rm FD}}\\
&  \equiv N_{\rm eff}^{\rm SM} + \Delta N_{\rm eff}\,.
\end{align}
 where $\rho_{\nu_i}$ is the energy density of $i^{\rm th}$ neutrino,
  $\rho_{\nu}^{\rm FD}$ is the energy density of a single neutrino species
  assuming a
  thermal Fermi-Dirac distribution and no energy gain during the electron-positron annihilation epoch, and
$\rho_{\rm DR}$ is the energy density of dark radiation. The theoretical
prediction in standard $\Lambda$CDM cosmology with only standard model
neutrinos contributing to $N_{\rm eff}$  is $N_{\rm eff}^{\rm SM}=3.046$; the extra
$0.046$ taking into account the energy gained by neutrinos during the
electron-positron annihilation
\cite{df1992,hm1995,dhs1997,fks1997,mmpp2002}. We have defined $\Delta
N_{\rm eff}$ as the  contribution by new physics.

\begin{figure}[!t]
	\begin{center}
		\includegraphics[width=0.44\textwidth]{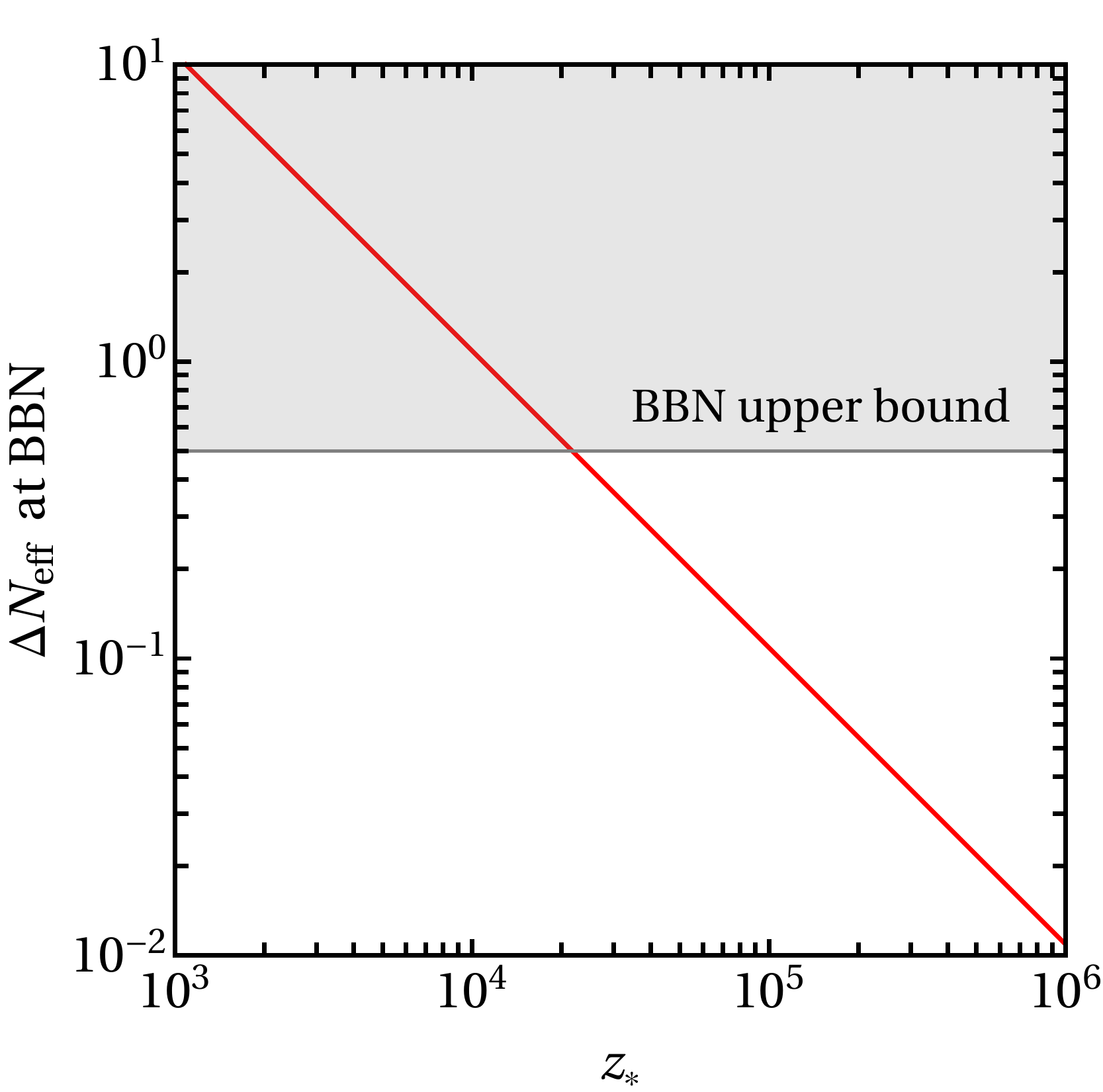}~\quad		\includegraphics[width=0.495\textwidth]{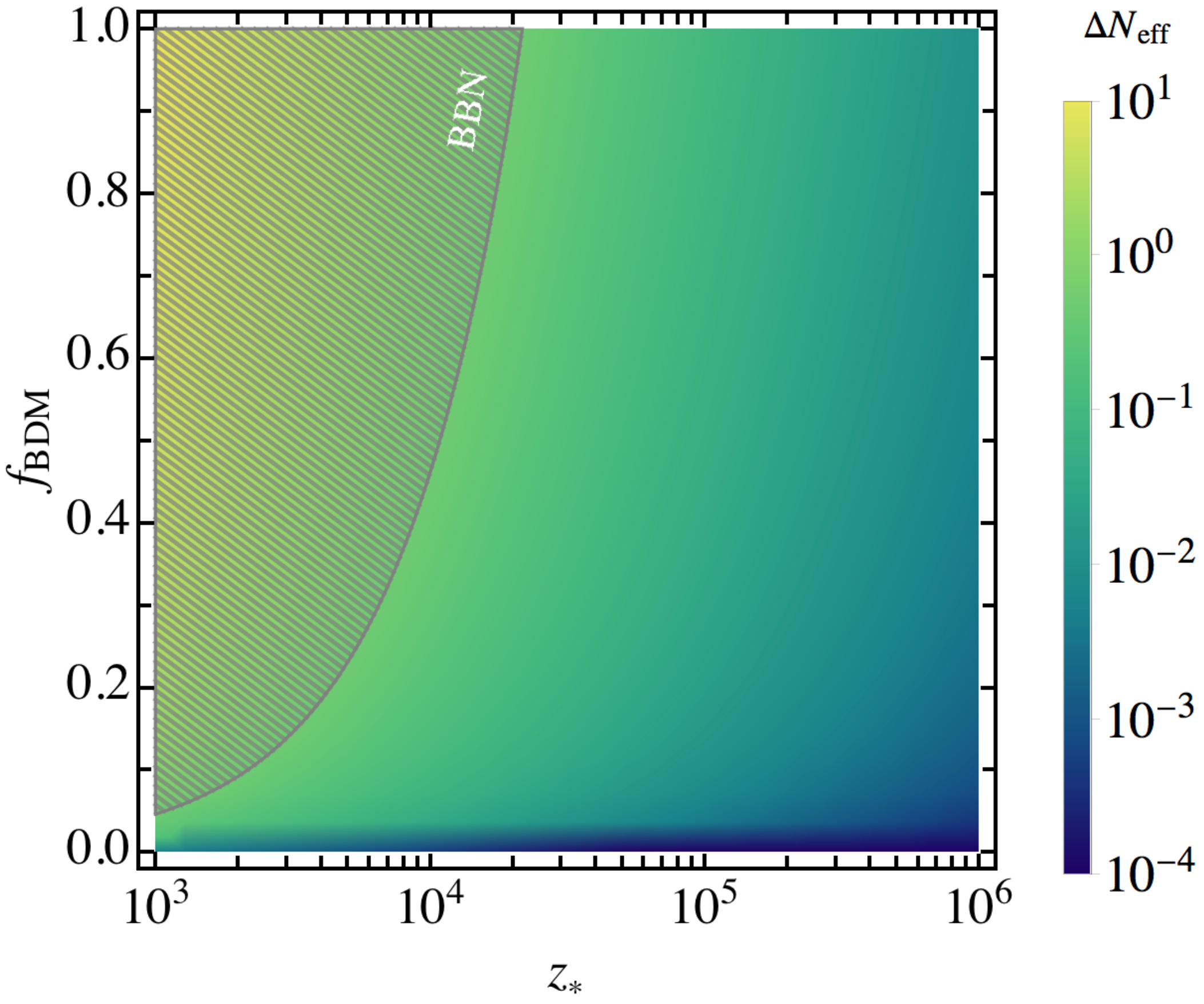}
		\caption{(Left) $\dNeff$ at BBN for a BDM transition at $\zs$ for $\fd=1$. The
                  upper bound is from primordial element abundance
                  measurement that restricts
                  $\Delta N_{\rm eff}<0.5$~\cite{2011PhLB..701..296M,
                    Cooke:2013cba}. (Right) The gray-striped region in the
                  $\zs-\fd$ plane is ruled out by the BBN constraint.\label{fig:bbn}}
	\end{center}
\end{figure}

The quantity $\Delta N_{\rm eff}$ is constrained by both CMB and big bang nucleosynthesis (BBN) observation data. 
The CMB experiments, such as Planck, are sensitive to the amount of radiation
present during recombination at $z\sim 1100$. The DR in our BDM model
 never thermalized with the visible sector. Also the time of phase transition
 must be 
 much earlier than the recombination era and the era of matter radiation
 equality in order to satisfy the current cosmological  constraints on the
 dark matter power spectrum. Therefore, the CMB anisotropy constraints on $N_{\rm eff}$ do not apply to
 our model because by the time of recombination BDM has the same background
 evolution as the CDM. Here we are implicitly assuming that all of the energy density in the 
BDM converts to dark matter. If this is not the case, and some energy remains as radiation, the constraint from CMB may also be important. 
However, if the phase transition happens after the BBN then
DR in BDM model will certainly contribute to the $N_{\rm eff}$ at the time of BBN and can be constrained
from the measurement of the primordial helium and deuterium
abundance~\cite{2011PhLB..701..296M, Cooke:2013cba}. The strongest BBN
constraints at present are given by $N_{\rm eff}=3.28\pm
0.28$~\cite{Cooke:2013cba} or $\Delta N_{\rm eff}\lesssim 0.5$. 

In Fig.\,\ref{fig:bbn}, we show the change in $\Delta N_{\rm eff}$ at the time of BBN as a function of $\zs$. For $\fd=1$, one finds $\zs\gtrsim2\times10^4$. Note the $\simeq 1/\zs$ scaling of the limit. This is obvious because the energy density due to BDM is fixed by requiring that it reproduce the present-day dark matter energy density. The excess radiation in the BBN epoch thus simply scales with the relative factor $(1+z_{\rm BBN})/(1+\zs)$. In the right panel of Fig.\,\ref{fig:bbn}, the variation of $\Delta N_{\rm eff}$ in the plane of $\zs-\fd$ is shown. The BBN constraint rules out the gray-striped region. However, we will see that the strongest effects of BDM would be on the matter  power spectrum, and through it also on the CMB anisotropy spectrum,
  resulting in  much stronger
  constraints than the BBN constraints on $\Delta N_{\rm eff}$. In other
  words, in the allowed parameter space for BDM, the extra contribution to
  the radiation energy density before the phase transition will  not be of concern.

\subsection{Matter Power Spectrum}

The main signature of BDM is through its impact on the  {matter density}
perturbations. Unlike in \lcdm cosmology, here the BDM can support waves
until the phase transition occurs at $\zs$, leading to acoustic
oscillations for  {$k$ modes inside the horizon at $\zs$.} We will also see that the nature of these acoustic oscillations in BDM is somewhat different from the acoustic damping seen in models where CDM is allowed to interact with a radiation like species.

\begin{figure}[t]
	\begin{center}
		\includegraphics[width=0.485\textwidth]{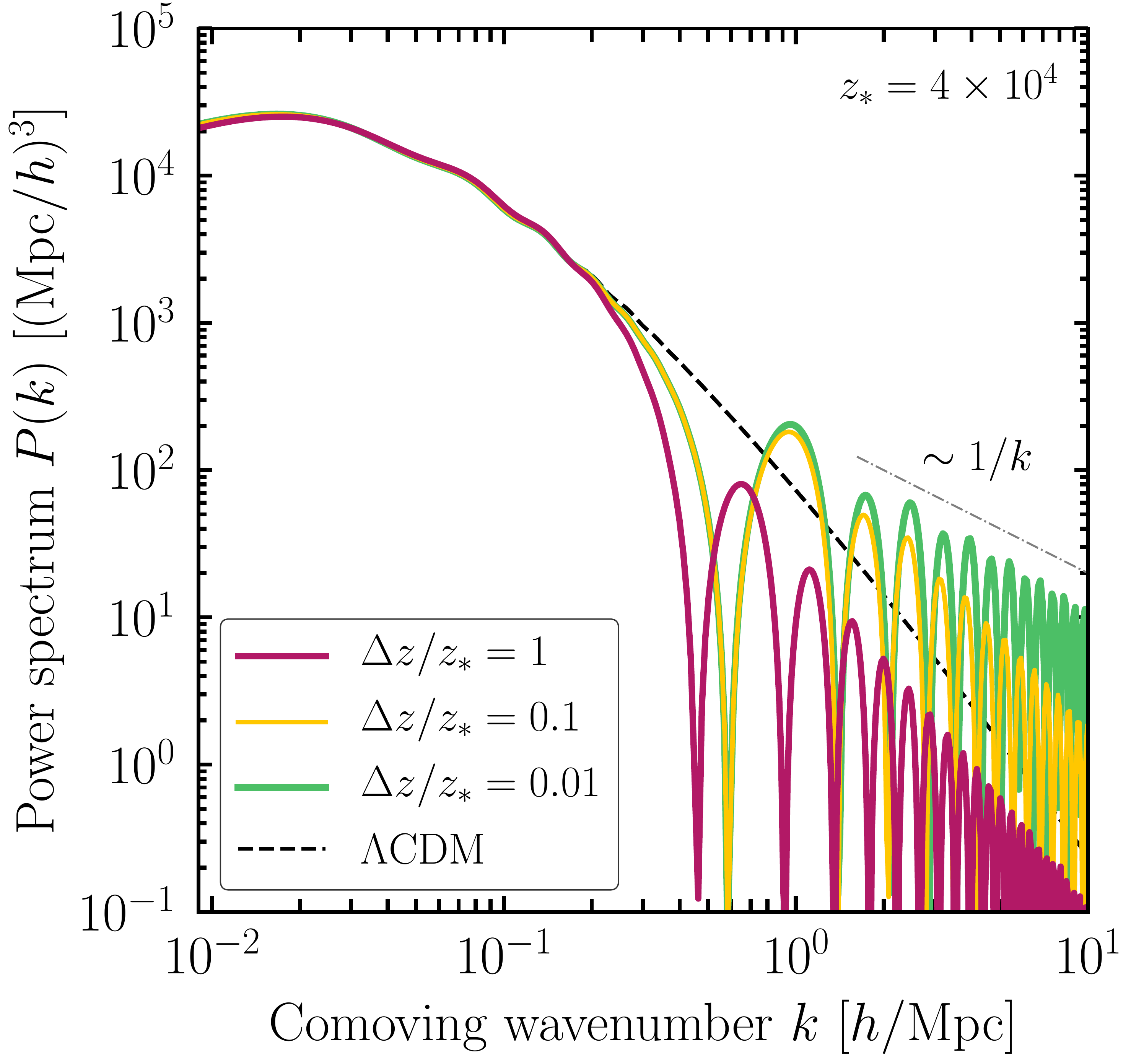}\,\,									\includegraphics[width=0.485\textwidth]{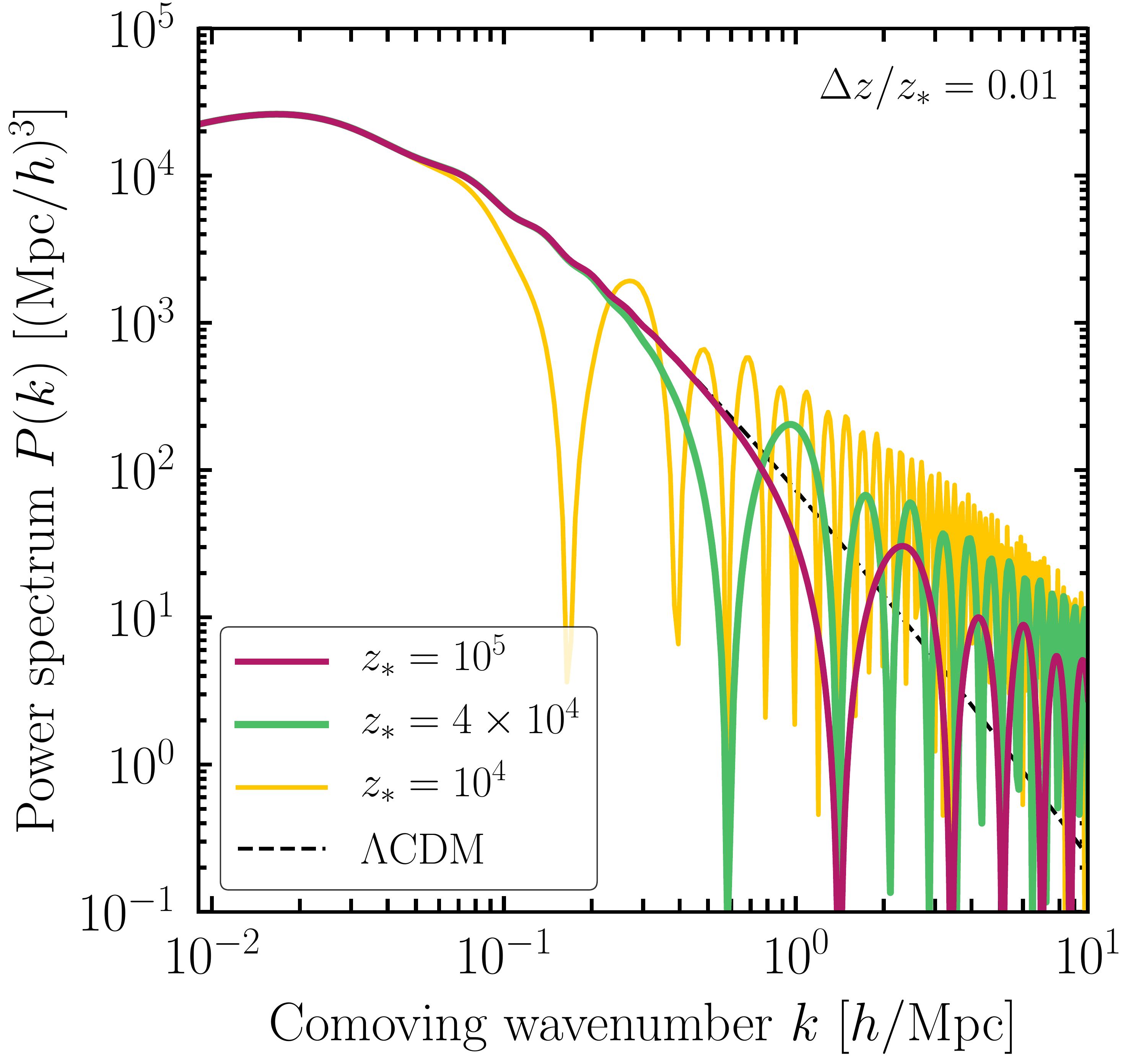}
		\caption{(Left) The matter power spectrum in a \lbdm cosmology with phase transition at $\zs=4\times 10^4$, but with 			different transition widths $\dz=1$ (purple), 0.1 (yellow), and 0.01 (green). (Right) The same matter power spectrum with 			different phase transition redshifts $\zs=10^5$ (purple), $4\times 10^4$ (green), and $10^4$ (yellow), but now for a fixed 			width $\dz=0.01$. For comparison, the \lcdm power spectrum is shown as a dashed black curve in both panels. Note the 			enhancement of power at the acoustic peaks at small scales relative to the \lcdm case.}
	\label{fig:mps}
	\end{center} 
\end{figure}

{In Fig.\,\ref{fig:mps}}, we show the dark matter power spectrum $P(k)$
in a \lbdm cosmology for different values of the additional parameters of
the theory, namely, the width of the phase transition $\dz$ and transition
epoch $\zs$. In the left panel, the green, yellow and purple curves
represent the matter power spectra for transition widths $\dz=0.01,~0.1$,
and 1, respectively. An important observation here is the relative
suppression of the spectrum for slower phase transitions. {This feature can
be attributed to the second term ($\propto \dot{\dd}$) on the LHS of
Eq.(\ref{eq:osc}). This term acts as a \emph{friction} term in the
oscillator equation and damps the fluctuations} during the span of the
phase transition. The effect of the phase transition epochs,
$\zs=10^4,~4\times 10^4$, and $10^5$, on the matter power spectrum is shown
in the right panel of Fig.\,\ref{fig:mps}. The value of $\zs$ decides the
scale {or wavenumber $\ks$} that was entering the horizon at the time of phase
transition. {All modes  with $k<\ks$} are unaffected and the power
spectrum is indistinguishable from the \lcdm case. {The modes with
  $k>\ks$} entered the horizon before the phase transition and experienced acoustic oscillations leading to the new features in the power spectrum. Of course, as one would expect, if the phase transition occurs at very early times the scale of acoustic oscillations moves to larger $k$, converging to the \lcdm model in the limit $\zs\to\infty$.

\subsubsection{Analytical Understanding of the Acoustic Peaks}

To understand the effect of the phase transition on the evolution of the
perturbations in a simple way, we first assume an instantaneous transition
at conformal time $\ts$, corresponding to the redshift $\zs$. Also, in this
section, we shall assume $z_*> z_{eq}$, i.e., the phase transition happens
inside the {radiation-dominated era. With this instantaneous phase
transition approximation, the evolution equation for $\dd$,
Eq. (\ref{eq:osc}), in each phase can be written as:}
\begin{align}
\ddot{\dd} + \dfrac{k^2}{3}\dd = - 4\ddot{\phi} - \dfrac{4}{3}k^2\psi \,, & \quad {\rm DR\ phase}\,,\label{eq:pert_2nd}\\
\ddot{\dd} + \dfrac{\dot{a}}{a}\dot{\dd} = -3\ddot{\phi} - 3\dfrac{\dot{a}}{a}\dot{\phi} - k^2\psi\,, & \quad {\rm DM\ phase}\,.\label{eq:pert_DM}
\end{align}
In this section, we shall be interested only in those modes which entered
the horizon much before $\ts$. We know that the metric perturbations decay
to zero after a mode $k$ enters the horizon during the radiation-domination
era. Therefore, if we ignore the potential-dependent source terms on the
RHS of Eqs. (\ref{eq:pert_2nd}) and (\ref{eq:pert_DM}), {the BDM
perturbation equation Eq. (\ref{eq:pert_2nd}) has an oscillatory solution}
\begin{equation}\label{eq:drsol}
	\dd(x<\xs) = A\cos x\,,
\end{equation} 
where we have defined the dimensionless quantity $x\equiv \kt/\sqrt{3}$ for
convenience. We kept only the cosine solution, {as required by the
  adiabatic  initial conditions} at $\tau=0$. This solution represents
perturbation modes with acoustic oscillations of frequency
$k/\sqrt{3}$. After the phase transition {the evolution of BDM is
  described by Eq. (\ref{eq:pert_DM}), yielding logarithmic growth during
  radiation domination,}
\begin{equation}\label{eq:dmsol}
	\dd(x>\xs) = B\ln x + C\,.
\end{equation}
{To fix the constants of integration, $B$ and $C$, we should know the 
$T^0_{\ 0}$ and $T^0_{\ i}$ components of the BDM stress-energy tensor at
the end of the phase transition or at the beginning of the DM phase.}
Because we are trying to study the cosmology in a model-independent way, we
consider the simplest possible choice, i.e., both these components of
$T^\mu_{\ \nu}$ are continuous during the phase transition. This assumption yields
\begin{equation}
 \begin{array}{rcl}
  (\dd)_{\rm DR} &=& (\dd)_{\rm DM}\,,\\[1ex]
  (\dot{\dd})_{\rm DR} &=& (\dot{\dd})_{\rm DM} - \dot{\phi}\,.
 \end{array}
\end{equation}
Since the potential $\phi$ decays after the mode enters the horizon, we
finally have continuous $\dd$ and $\dot{\dd}$ across the phase transition
happening $x=\xs$. Their values at $\xs$ act as the initial conditions for
the perturbations in the ensuing DM phase. Therefore, the solutions of
Eq.(\ref{eq:drsol}) and (\ref{eq:dmsol}) need to be matched at $x=\xs$ by
equating $\dd$ and $\dot{\dd}$.  The final DM phase solution is then given by
\begin{equation}
\begin{array}{l}
\dd(x>\xs) = A\cos\xs - A\xs\sin\xs\ln\left(\dfrac{x}{\xs}\right)\,.
\end{array}
\label{eq:gensoln}
\end{equation}
 The constant $A$ is set by the {initial conditions or initial curvature
   perturbation.} Although the evolution is always logarithmic {during
   radiation dominated era}, depending on the value of $\xs=\kt_*/\sqrt{3}$, two extreme cases are possible in the DM phase:
\begin{align}\label{eq:caseI}
\dd(x) &= (-1)^nA\,, \quad\quad\quad\quad\quad\quad\quad\quad{\rm if}~\xs = n\pi\,,\vspace{0.28cm}\\
\dd(x) &= (-1)^{n+1}A\xs\ln\left(\dfrac{x}{\xs}\right)\,,\quad\quad{\rm if}~\xs = \left(n+\frac{1}{2}\right)\pi\,.\label{eq:caseII}
\end{align}
Here $n$ is any integer. For the modes with $k$ such that $\xs=n\pi$, the
density fluctuation does not grow at all after the phase
transition. Eventually these {modes at even multiples of $\pi/2$ (or
  zeros of the sine) at the
  phase transition} will carry less power and
correspond to the minima in the power spectrum. On the contrary, if
$\xs=(n+1/2)\pi$, these {modes at odd multiples of $\pi/2$ at the phase
  transition (extrema of the sine function) will have logarithmic growth
with maximum  slope. This large initial slope or prefactor is responsible
for fast initial growth which may, for sharp phase transitions,
overtake the \lcdm perturbation giving acoustic peaks that overshoot the
\lcdm power for the same $k$ modes.}

Physically these two families of solutions are caused by the different
\emph{velocities} of the perturbation at $\ts$. The modes which were crossing zero
and had maximum velocity at $\xs$, i.e., $|\dot{\dd}(\xs)|=A$, will
{continue moving \emph{ballistically} with the same bulk velocity in the collisionless DM phase, until the
  initial velocity is redshifted away. This inherited  extra \emph{bulk
    velocity kick} w.r.t what we expect from just gravitational infall in
  standard CDM, results in a faster logarithmic growth for these modes}
compared to all other modes. On the other hand, the modes having maximum
displacement and zero velocity at $\xs$, i.e., $|\dot{\dd}(\xs)|=0$, will
not grow initially because the prefactor of the logarithmic term in Eq.\,(\ref{eq:gensoln})  vanishes. All other modes which do not belong to these two
extreme cases also have logarithmic growth but with relatively smaller
slope. After $\tau=\teq$ in the matter-dominated era, all modes grow as
$\dd\sim a \sim \tau^2$. These different types of mode evolution will
reflect themselves in the shape of the matter power spectrum. {In particular
it is the peculiar or bulk velocities of acoustic oscillations, and hence the sine mode, which
get imprinted in the matter power spectrum, similar to the phase shift
experienced by the baryon acoustic oscillations w.r.t. to the acoustic
oscillations imprinted in the CMB \cite{sz1970,eh1998}.}

\begin{figure}[t]
	\begin{center}
	\includegraphics[width=0.5\textwidth]{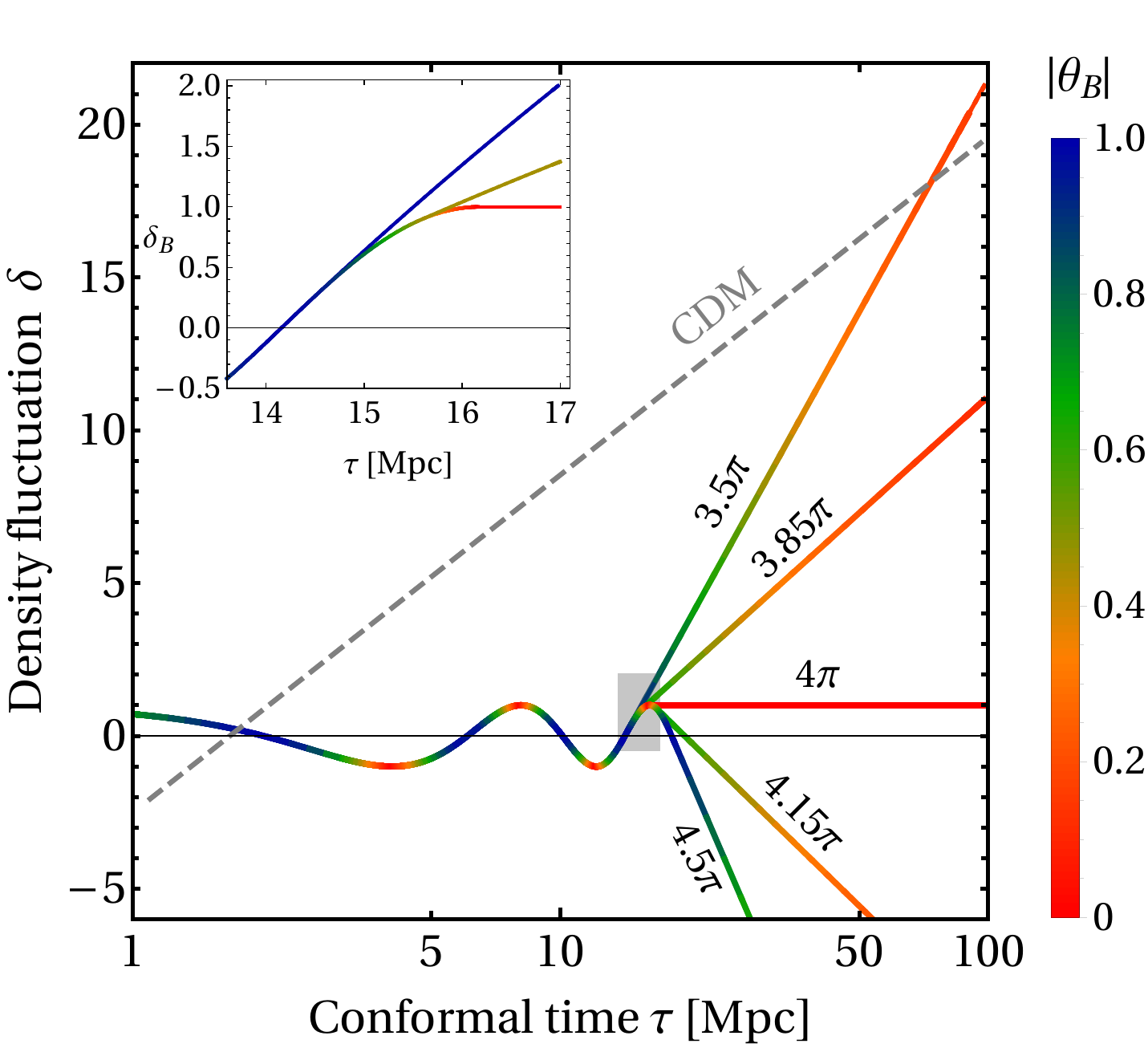}\,\,
		\includegraphics[width=0.46\textwidth]{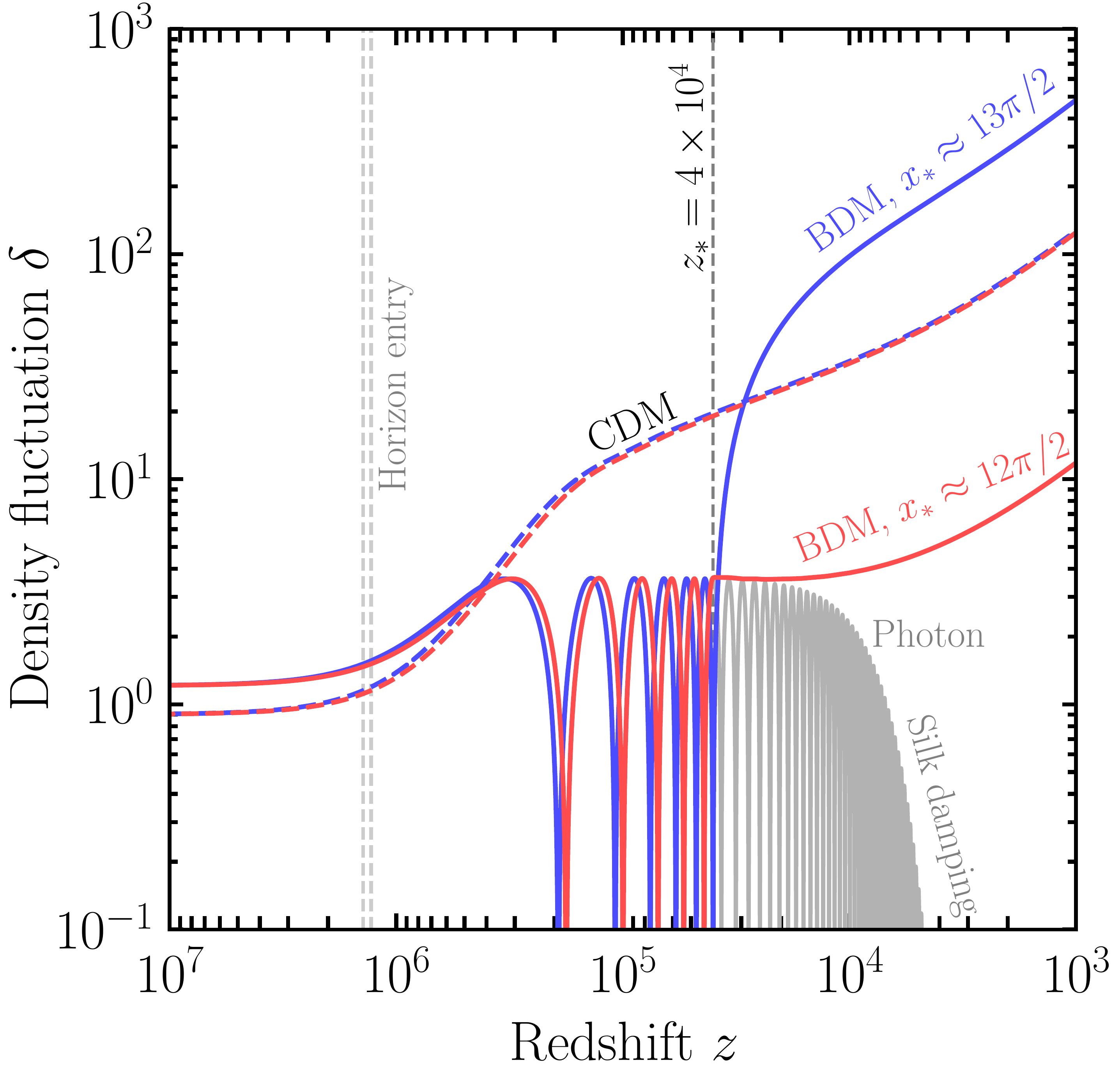}
		\caption{(Left) Analytical solution of the BDM linear perturbation equations for the mode $k=2h$/Mpc. Five curves are shown with different $\ts$s corresponding to $\xs=3.5\pi,~3.85\pi,~4\pi,~4.15\pi$ and $4.5\pi$, respectively. The colour of the curves represent the absolute value of the velocity perturbation $|\td|$. A zoomed-in version of the gray region is shown in the inset. The dashed, gray line shows the evolution of the same mode in CDM perturbation. (Right) Numerical solution for evolution of two modes of BDM perturbation $\dd$ corresponding to a maximum ($\xs\approx 13\pi/2$) and a minimum ($\xs\approx 12\pi/2$) in the matter power spectrum. Corresponding CDM mode evolutions in \lcdm cosmology are also shown as dashed curves. The other parameter values are $\zs=4\times 10^4, ~\dz=10^{-2}$ and $\fd=1$.}\label{fig:mode_evolve}
	\end{center}
\end{figure}

In the left panel of Fig.\,\ref{fig:mode_evolve}, we show the evolution of
$\dd$ as a function of $\tau$ for the mode $k=2h$/Mpc for five different
values of $\ts$. This is simply the analytical solution shown in
Eq.\,(\ref{eq:gensoln}). The color-coding represents the absolute value of
$\dot{\dd}$, hence the absolute value of $\td$. The transition epochs are
chosen such that $\xs=3.5\pi,~3.85\pi,~4\pi,~4.15\pi$ and $4.5\pi$,
respectively. Until the phase transition the evolution is identical, but
depending on $\xs$ the curves emanate from the phase transition point with
different colours (i.e., velocities) which can be seen in the zoomed-in
version of the gray region in the inset. As was argued in
Eq.(\ref{eq:caseII}), the cases $\xs=3.5\pi$ and $4.5\pi$ correspond to the
{extrema of the sine function  (or the peculiar velocities at the
phase transition)  which show fast growth} of the perturbations, resulting
in excess power at the acoustic peaks seen in Fig.\,\ref{fig:mps}. The case
with $\xs=4\pi$ is {the zero of the sine function} and has zero velocity
but maximum {density perturbation} at $\tau=\ts$ and remains frozen at
this value, lagging behind all other modes at late times. They correspond
to the dips in the oscillatory part of the power spectrum. Other cases of
$\xs=3.85\pi$ and $4.15\pi$ have intermediate velocities at $\ts$. Note how
all peculiar velocities redshift as $\sim 1/a$ after the phase
  transition. Eventually, of course, the peculiar velocities sourced by gravitational potentials
  will take over.

In the right panel of Fig.\,\ref{fig:mode_evolve}, we see that the
numerical results show the same behavior as above. The two modes,
$k=4.305h$/Mpc and $4.661h$/Mpc, roughly correspond to $\xs\simeq 12\pi/2$
and $13\pi/2$, respectively, for a phase transition at $\zs=4\times
10^4$. These modes lead to a dip and a peak, respectively, in the matter
power spectrum. The perturbations remain constant at their initial values
until the time of their respective horizon entry which happens when
$\kt\simeq 1$. Afterwards they start oscillating with a frequency
$k/\sqrt{3}$. They continue to oscillate until $\ts$, thereafter they start
growing as $\sim \ln\tau$ during the radiation-domination era and as $\sim
\tau^2$ in the matter-domination era. The same modes for $\dc$ in a \lcdm
cosmology are also shown in the dashed curves. As discussed in the
preceding paragraph, the modes starting with extra \emph{bulk velocity
  kicks} from the pre-phase transition oscillations overshoot the \lcdm
value and eventually acquire more power. They give rise to the peaks in
Fig.\,\ref{fig:mps}. Whereas those perturbations which were at their
maximum values at the time of phase transition (hence, zero velocity) grow
at a much slower rate and lead to the dips in Fig.\,\ref{fig:mps}. Indeed
the  mode labeled by $\xs\approx13\pi/2$ (solid blue, {peak of the sine
function}), grows faster and goes above the \lcdm curve (dashed blue, {zero
of the sine function}), while the  mode labeled by $\xs\approx12\pi/2$ (red curve) remains below it. This gives rise to the oscillatory feature in the matter power spectrum in Fig.\,\ref{fig:intro_fig}, with the upper envelop of the oscillations going above the \lcdm expectation.

From  Eq.(\ref{eq:caseII}), we note that the absolute value of maximum perturbation is proportional to the wavenumber $k$. Hence the transfer function $T(k)_{\rm max} \sim k$. As a result, we expect the envelop of the peaks of the $P(k)$ to scale as $\sim 1/k$,
\begin{equation}\label{eq:maxima_scaling}
P(k)_{\rm max} \equiv \dfrac{2\pi^2\mathcal{P}_p}{k^3}T(k)_{\rm max}^2 \sim 1/k\,,
\end{equation}
where $\mathcal{P}_p$ is the primordial scalar power spectrum defined as
\begin{equation}
 \mathcal{P}_p = A_s\left(\frac{k}{k_0}\right)^{n_s-1}\,,
\end{equation}
 in terms of the amplitude $A_s$, the scalar index $n_s=0.96$, and the pivot
 scale $k_0$~\cite{Ade:2015xua}. The $1/k$ upper envelop of $P(k)$ predicted by
 Eq.(\ref{eq:maxima_scaling}) is evident  in
 Fig.\,\ref{fig:mps} {for the case of fast transition, $\dz=0.01$, as shown by the dashed gray line.}


\subsubsection{Odd-Even Peak Asymmetry}
\begin{figure}
	\begin{center}
		\includegraphics[width=\textwidth]{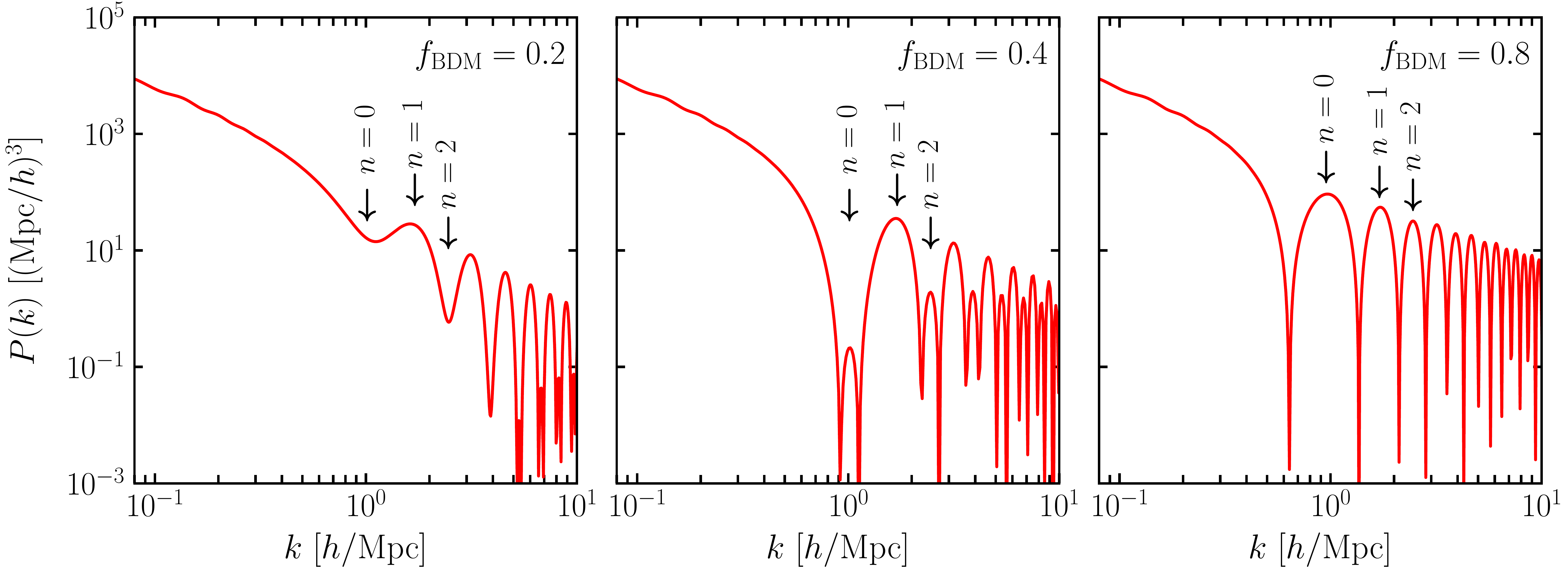}
		\caption{The matter power spectra for three different BDM
                  fractions $\fd= 0.2,~0.4$ and 0.8. The first three
                  `peaks' numbered by $n=0,1,2$ ({see Eq. (\ref{eq:caseII})} for details) are marked with black arrows.}\label{fig:odd_even1}
	\end{center}
\end{figure}

\begin{figure}[t]
	\begin{center}
		\includegraphics[width=0.49\textwidth]{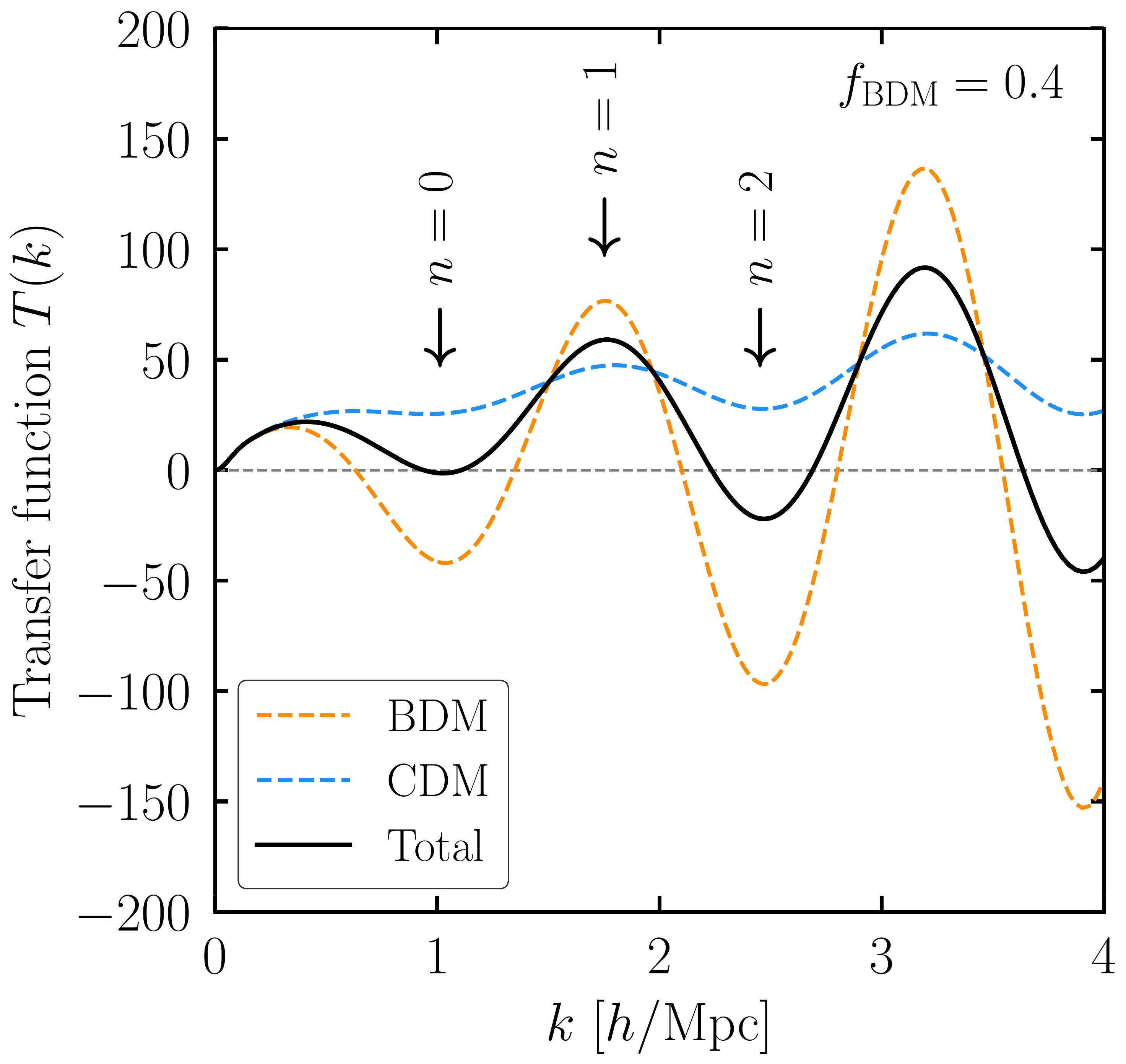}~\,\  \includegraphics[width=0.49\textwidth]{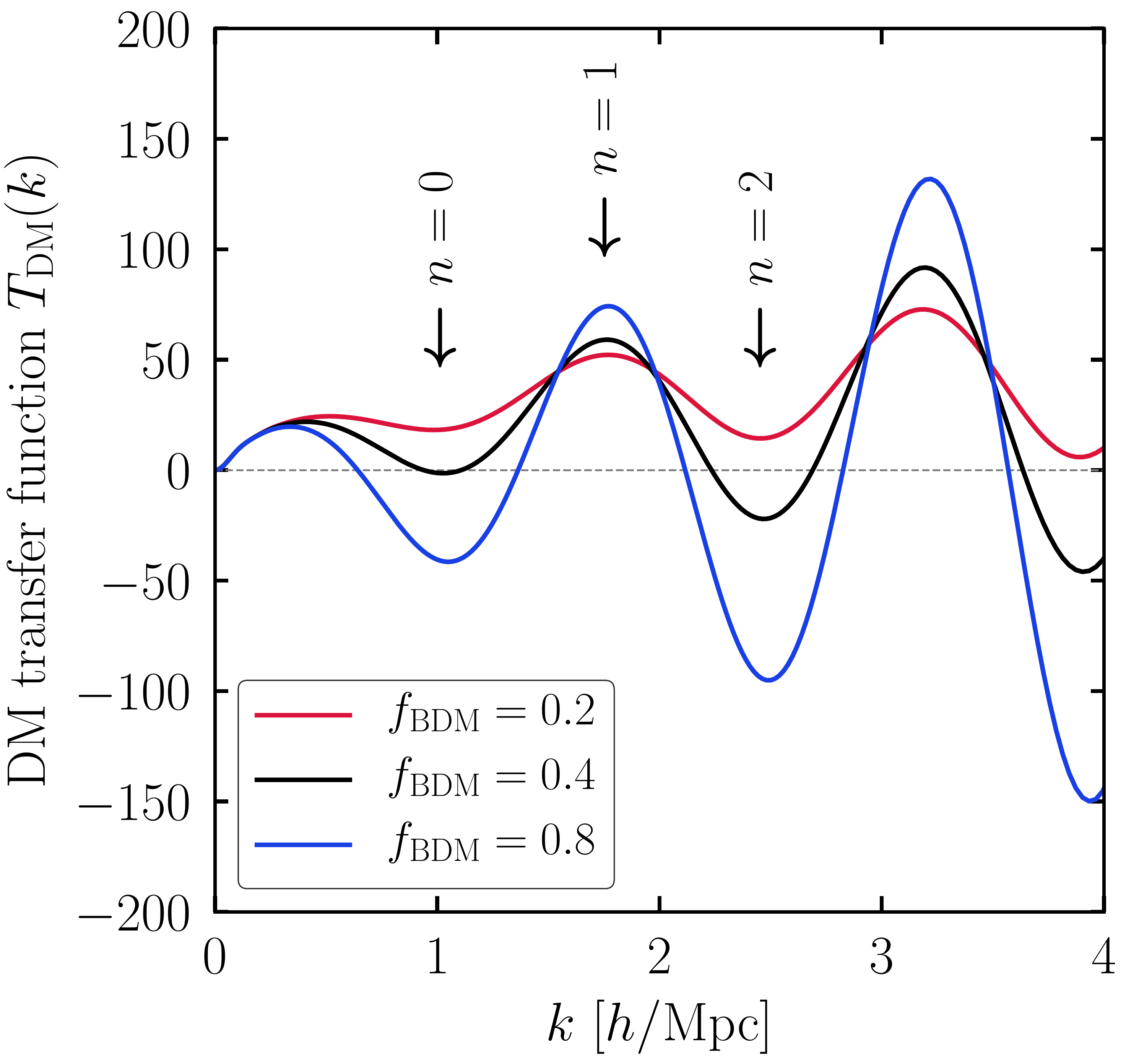}
		\caption{(Left) Individual transfer functions of BDM $\tbdm(k)$ (dashed orange), CDM $\tcdm(k)$ (dashed light blue), and the total dark matter transfer function $\tdm=\fd\tbdm+(1-\fd)\tcdm$ (solid black) at redshift $z=3000$ for $\fd=0.4$. (Right) Comparison between dark matter transfer functions $\tdm(k)$ computed for three different fractions of BDM, viz., $\fd=0.2~{\rm (red)},~0.4~{\rm (blue)}$, and $0.8~{\rm (green)}$. Other phase transition parameters are $\zs=4\times 10^4,\ \dz=10^{-2}$.}\label{fig:odd_even2}
	\end{center}
\end{figure}

An interesting asymmetry in the heights of the power spectrum peaks becomes
apparent if the fraction of BDM is neither 0 nor 1. Three representative
cases are shown in Fig.\,\ref{fig:odd_even1} with $\fd=0.2,~0.4$ and 0.8,
respectively. {We have numbered the peaks according of value of $n$ in Eq. (\ref{eq:caseII}) with the first peak given by $n=0$ corresponding to first zero
  crossing of cosine (density) or first extrema of sine (peculiar velocity).} First, we concentrate on the $\fd=0.4$ case and observe that the heights of the odd-numbered peaks are greater than the even-numbered ones. To understand the reason behind this asymmetry, we plot in the left panel of  Fig.\,\ref{fig:odd_even2} the individual BDM and CDM transfer functions $\tbdm$ (dashed orange) and $\tcdm$ (dashed light blue), along with the total dark matter transfer function $\tdm(k)$ (solid black), which is defined as
\begin{equation}
	\tdm(k)=\fd\tbdm(k)+(1-\fd)\tcdm(k)\,,
\end{equation}
at a redshift $z=3000$. Oscillations with amplitude growing with $k$ are
present in the BDM transfer function, $\tbdm(k)$, as expected. However, we
now see that such oscillations, albeit with smaller amplitude, are also
imprinted in the CDM transfer function $\tcdm$. This is {the result of the
  CDM responding to the gravity of BDM or the gravitational potential
  $\phi$ which} has contribution from $\tbdm$. Further, we observe the
relative sign between the two transfer functions. At the positions of the
peaks of  $\tbdm$, the two transfer functions have the same sign and
reinforce each other resulting in a larger magnitude of $\tdm$. On the
other hand, at the troughs of $\tbdm$ they have opposite signs and can
partially cancel each other. These shallower troughs of the $\tbdm$, that
are below zero, appear as smaller peaks in the matter power spectrum. This
leads to the asymmetry between the consecutive maxima in the matter power
spectrum in Fig.\,\ref{fig:odd_even1}. {Physically, the initial
  velocities of BDM at the phase transition for $k$ modes corresponding to even values of $n$ were such that the (BDM) matter
  flowed from out of the initial overdensities, which were the same for CDM
  and BDM according to the adiabatic initial conditions, and flowed into
  the initial underdensities and thus reducing the amplitude of perturbations for
  those modes. For the modes corresponding to odd values of $n$, the BDM matter
  flowed into the CDM overdensities and out of the CDM underdensities,
   increasing the density contrast.}

Both components of the dark matter are needed in sizeable amount for the
odd-even acoustic peak asymmetry to be prominent. This is evident from the
$\fd=0.2$ and 0.8 plots in Fig.\,\ref{fig:odd_even1} and {the right panel of
Fig. \ref{fig:odd_even2}. For $\fd=0.2$,
  the BDM has a sub-dominant contribution to the total power spectrum
  and the out of phase extrema of BDM (even-$n$) only result in giving
   minima in the total power spectrum. Thus only the odd-$n$ modes result
   in acoustic peaks in the total matter power spectrum. As we increase
   $\fd$, the minima in the total transfer function  become deeper
   and deeper and at some point cross zero (see Fig. \ref{fig:odd_even2}, right panel). Once the total transfer
   function has a zero crossing,  the zero-crossings become the deep minima in
   the matter power spectrum and the minima of the transfer function appear
   as additional peaks, doubling the number of acoustic peaks in the total
   power spectrum.}  
\begin{figure}[t]
 \begin{center}
  \includegraphics[width=0.7\textwidth]{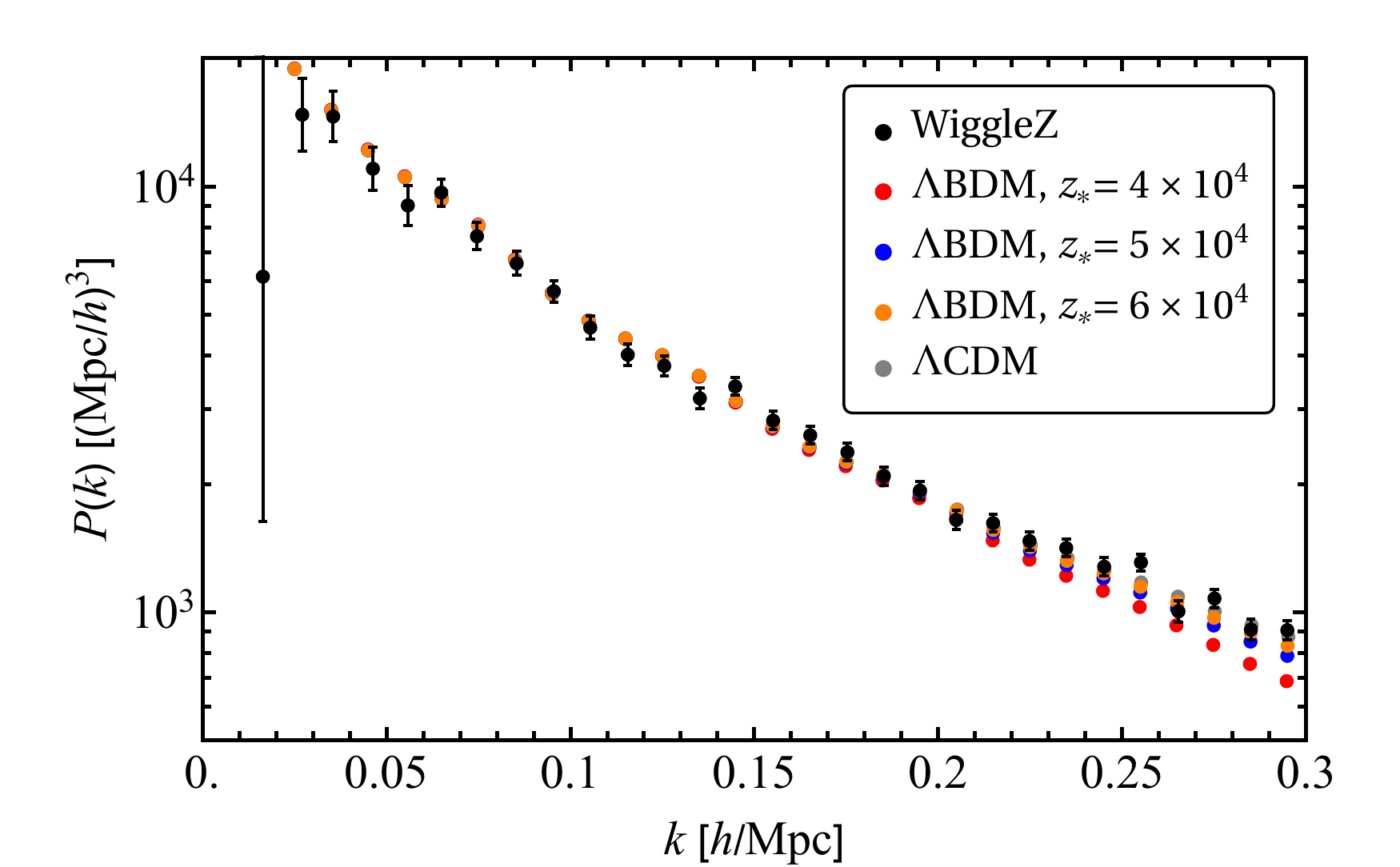}
  \caption{Comparison between the observed matter power spectrum by WiggleZ~\cite{2012PhRvD..86j3518P} and the theoretical power spectra for different transition redshifts, $\zs=4\times 10^4,~5\times 10^4,~6\times 10^4$.}\label{fig:wigglez}
 \end{center}
\end{figure}

The asymmetry, i.e., the relative heights of the consecutive maxima in the
power spectrum {is fixed 
once the initial peculiar velocities have redshifted away. Subsequently,
the BDM and CDM can be treated as a single collisionless cold fluid,
with a modified power spectrum which grows linearly with redshift
identically to the CDM fluid in the standard \lcdm cosmology. In
particular, subsequent linear growth does not change the shape of the power
spectrum and the asymmetry and acoustic features persist until today in
linear theory.}

\subsection{Qualitative Constraints from the Matter and CMB Power Spectrum}

As we have already discussed, {the effective relativistic degrees of
  freedom during BBN already gives interesting constraints on the redshift
  of phase transition. A} more stringent lower bound on $\zs$ is given by
the measurement of the dark matter power spectrum. From Fig.\,\ref{fig:mps}
we see that even a value $\zs=10^4$  predicts a sharp drop in power at
$k=0.1h$/Mpc near the second BAO peak. {These scales are well measured at
many redshifts by the current galaxy surveys like SDSS
\cite{Percival:2006gt} and WiggleZ \cite{2012PhRvD..86j3518P} and therefore
$\zs=10^4$ is clearly ruled out by the current matter power spectrum
measurements. We show the  WiggleZ data from the redshift range $0.5<z<0.7$
and theoretical \lbdm power spectrum  using the flat \lcdm best-fit model
parameters in the Table VII of Ref.\,\cite{2012PhRvD..86j3518P}  in
Fig.\,\ref{fig:wigglez} for different $\zs$. We have used the same binning as the WiggleZ data for the theoretical power spectrum and convolved it with the WiggleZ window function. As we can see,  even
when restricting to approximately linear modes, $k<0.3h$/Mpc, we can
already rule out $\zs$ smaller than $\sim 5\times 10^4$ by eye. We remind
that this is a crude estimate, and to be more accurate one needs to do a
more detailed study with degeneracies with other \lcdm parameters, such as $n_s$, taken into account. }

\begin{figure}[t]
	\begin{center}
		\includegraphics[width=0.485\textwidth]{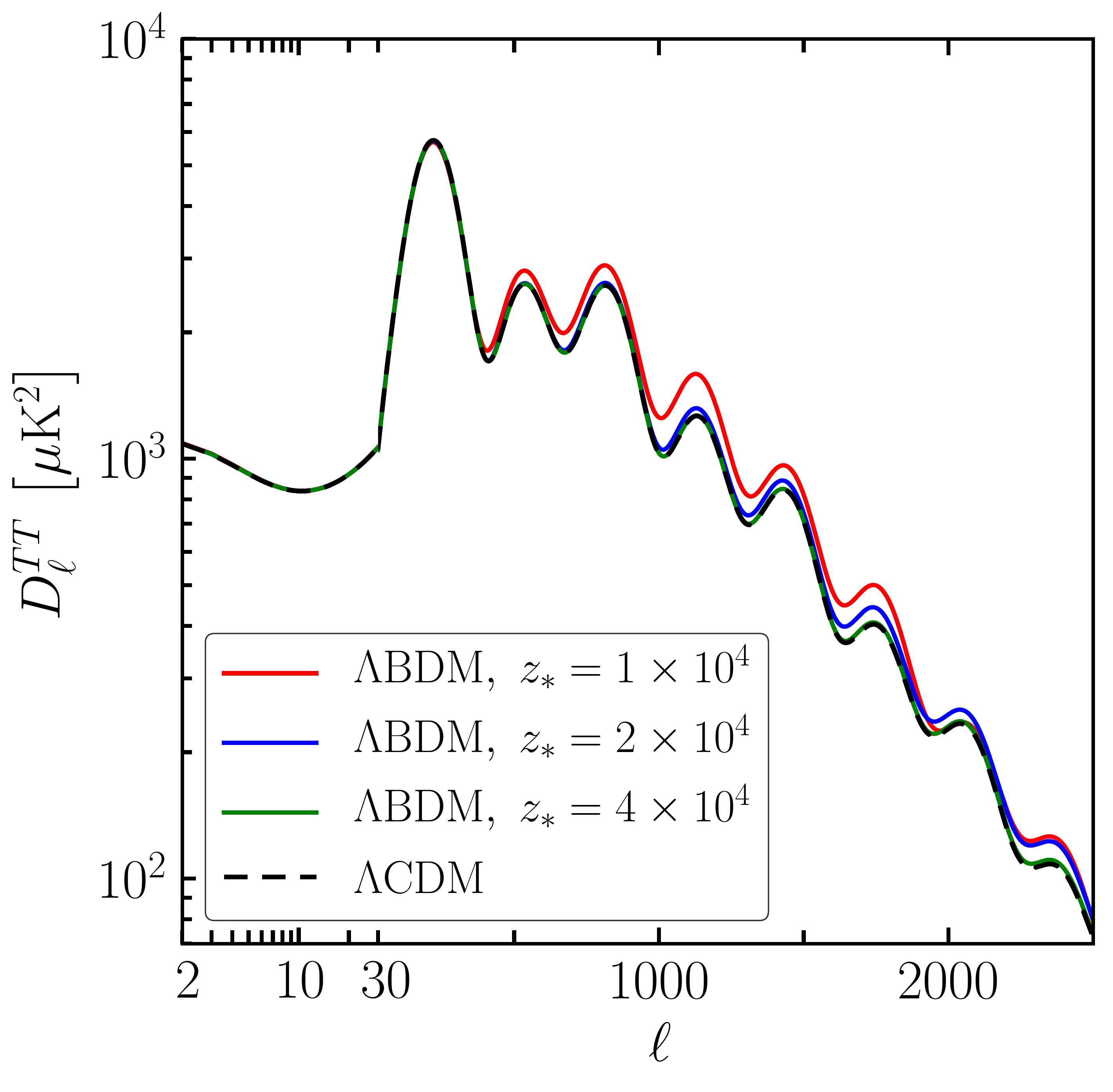}\ \ \,
		\includegraphics[width=0.485\textwidth]{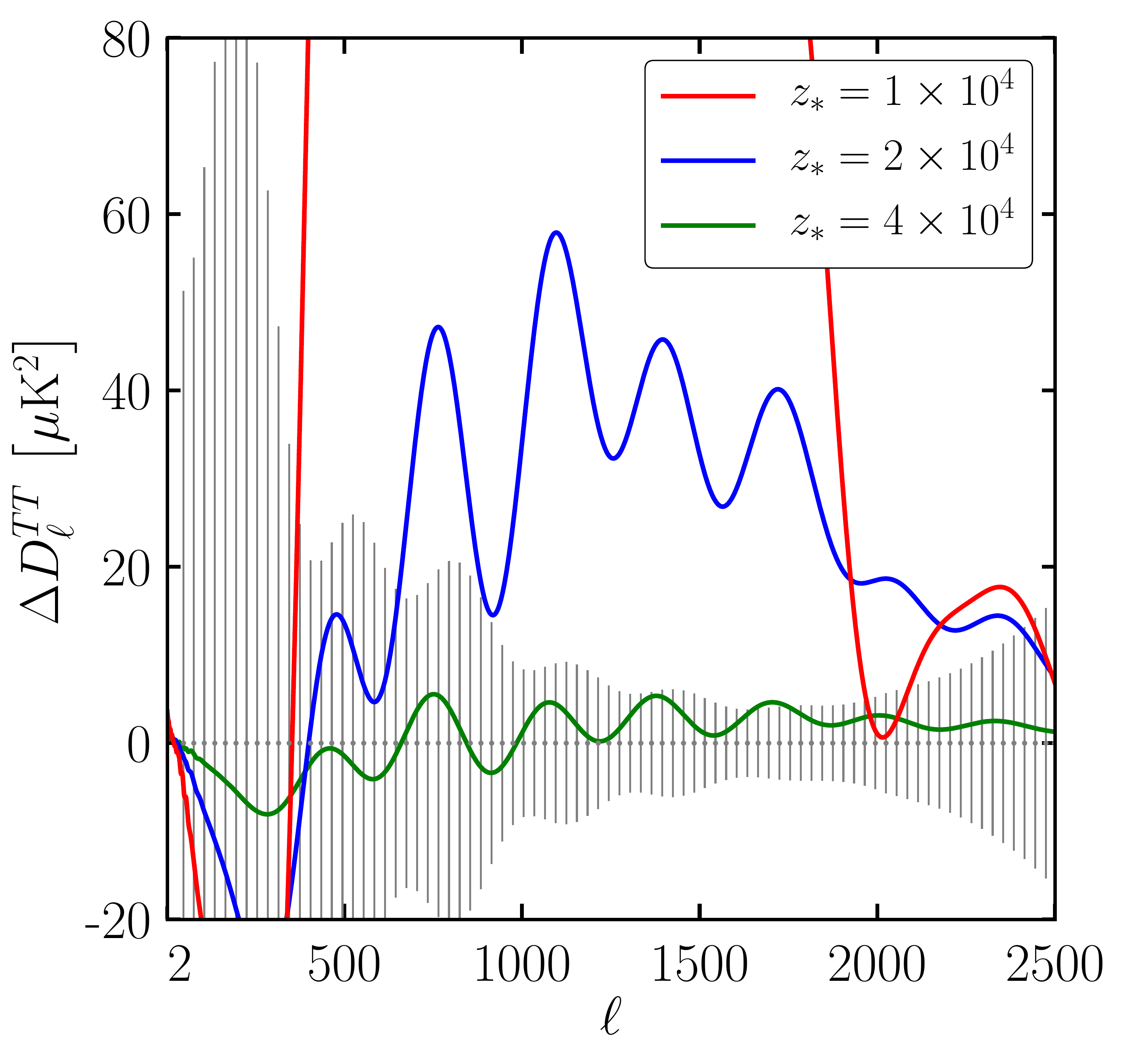}
		\caption{(Left) Effects in the CMB TT power spectrum is shown for $\zs=10^4$ (red), $2\times 10^4$ (blue), and $4\times 10^4$ (green) with $\dz=0.01$ and $\fd=1$. (Right) The difference between \lbdm and \lcdm powers $\Delta D^{TT}_\ell \equiv (D^{TT}_\ell)_{\Lambda{\rm BDM}}-(D^{TT}_\ell)_{\Lambda{\rm CDM}}$. We compare this $\Delta D^{TT}_\ell$ with the Planck high-$\ell$ ($47\leq\ell\leq 2499$) binned data, and conclude that the Planck data puts a rough lower limit on the transition redshift $\zs\gtrsim 4\times 10^4$.}\label{fig:cmb}
	\end{center}
\end{figure}

{We will also expect modifications to the  CMB anisotropy power spectrum at
small angular scales as it is sensitive to the total dark matter power
spectrum at the time of recombination.} In a flat Universe, the mode $\ks$
corresponds to an approximate angular scale of $\ell_*\simeq \ks\tau_0$
where $\tau_0=1.4\times 10^4$ Mpc is the conformal time today. Therefore
the observability of this effect in the CMB angular power spectrum would
depend on the value of $\ks$. The smallest scale probed by the current
Planck experiment corresponds to $\ell_{max}=2500$ implying a value of
$\ks\simeq 0.2h$/Mpc, therefore a sensitivity to $\zs \lesssim 6\times 10^4$. The typical changes expected in the CMB TT angular
power spectrum are shown in the left panel of Fig.\,\ref{fig:cmb} for
$\zs=10^4$ (red), $2\times 10^4$ (blue) and $4\times 10^4$ (green). {We use}
the best-fit values of \lcdm parameters from the Planck
experiment~\cite{Ade:2015xua}. In the right panel of Fig.\,\ref{fig:cmb},
we plot the difference between \lbdm and \lcdm powers $\Delta D^{TT}_\ell
\equiv (D^{TT}_\ell)_{\Lambda{\rm BDM}}-(D^{TT}_\ell)_{\Lambda{\rm CDM}}$
together with the Planck high-$\ell$ ($47\leq\ell\leq 2499$) binned data
error bars. We conclude that values of transition redshift
$\zs\lesssim 4\times 10^4$ are not consistent with the CMB data.  The
next-generation CMB observation experiments promise to probe even smaller
angular scales and correspondingly smaller $k$
values~\cite{Abazajian:2016yjj}. 

{The matter power spectrum and the CMB power spectrum at present would give
  comparable constraints on the BDM parameters ($\zs, \dz, \fd$) with the constraints
  from the matter power spectrum expected to be stronger. We leave a more
  detailed Markov Chain Monte Carlo study of the \lbdm parameters using
  current CMB temperature and polarization data and matter power spectrum
  for a future publication.}

\section{Summary \emph{\&} Conclusions}
{We have studied the cosmological consequences of a class of dark matter
models defined by two main properties: 
\begin{enumerate}
	\item The time when the dark matter
	becomes non-relativistic coincides with it also becoming
	collisionless and the dark fluid is strongly interacting before this phase transition.
	\item This phase transition happens much later than the decoupling
	of dark matter from the visible sector and in particular happens after BBN
	and before recombination.
\end{enumerate}
Before the phase transition to
 non-relativistic collisionless dark matter, the radiation-like particles
were tightly coupled together and constituted a perfect fluid. The pressure
in the fluid supports acoustic oscillations and stalls the growth of
density perturbations during the period between a mode's horizon entry and
the phase transition at $\ts$. The consequence of the above two features is that
 the non-relativistic phase of the dark matter starts with a non-zero
 peculiar velocities which are a sinusoidally oscillating function of the
 mode $k$, and which are
 90 degrees out of phase w.r.t. the density fluctuations, inherited from
 the previous tightly coupled relativistic phase. The initial evolution in
 the collisionless phase is ballistic until the initial acoustic peculiar
 velocities have been redshifted away, hence the name Ballistic Dark Matter
 or BDM. }  Afterwards the perturbations grow in a similar fashion as in
the \lcdm cosmology. {The initial evolution after the phase transition
  of a mode is therefore driven almost entirely by the peculiar velocities
  at the phase transition. The modes which had the maximum velocity at
$\ts$ grow  fastest and the modes for which the density perturbation was
at the maximum amplitude and hence had zero velocity have the slowest
initial growth. The acoustic oscillations before the phase transition are
thus imprinted on the dark matter power spectrum. For fast phase
transitions, the acoustic peaks in the matter power spectrum, driven by
high initial peculiar velocities, can exceed the
\lcdm power.}  The excess growth of power relative to the \lcdm case can be
suppressed if the phase transition happens rather slowly. A gradual
variation of the EoS of the dark sector fluid leads to damping of perturbations. 

{If BDM does not dominate the matter energy density in the Universe then an
asymmetry arises in the peak heights of the matter power spectrum. This
happens because the transfer functions of CDM and BDM can be in-phase or
out-of phase at the extrema of the BDM transfer function. The minima
and maxima of the BDM transfer function have opposite sign and would
give rise to similar amplitude acoustic peaks if BDM formed all of dark
matter. The CDM transfer function on the other hand does not change sign as a
function of $k$. Therefore successive extrema of the BDM would have
alternatively the same and the opposite sign to that of CDM and the two can add
constructively or destructively. The acoustic peaks in the total matter
power spectrum would be therefore alternate between enhancement and
suppression giving an odd-even peak asymmetry. }

{By varying the three parameters of our BDM model, the redshift of phase
  transition, $\zs$, the duration of phase transition, $\dz$, and the
  fraction of dark matter formed by BDM, $\fd$, we can get a rich variety
  of features and, in particular, tune the matter power spectrum to be
  enhanced or suppressed at particular wavenumbers $k$. We have shown, by
  comparison with existing data, that for fast transitions and all of DM
  formed by BDM, the phase transitions must happen at $\zs>5\times
  10^4$. Our results indicate that Ballistic Dark Matter has rich
  cosmological phenomenology and motivate a more detailed study of the consequences of
  such a dark matter model on the large scale structure, in particular in
  the non-linear regime, in the future.}

\begin{acknowledgments}
	We gratefully acknowledge useful discussions with Subhajit Ghosh,
        Sourendu Gupta, and Nilmani Mathur. BD was partially supported
        through a Ramanujan Fellowship of the Dept. of Science and
        Technology, Government of India and the Max-Planck-Gesellschaft funded 
        Max Planck Partner Group between Max Planck Institute for Physics, Munich
        and Tata Institute of Fundamental Research. RK was 
        supported by Science and Engineering  Research Board (SERB) of  
        Department of Science and Technology, Govt. of India through SERB grant 
        ECR/2015/000078 and Max Planck Partner Group between Max Planck Institute for
        Astrophysics, Garching  and Tata Institute of Fundamental Research. 
\end{acknowledgments}

\bibliographystyle{jhep}
\bibliography{bdm}

\providecommand{\href}[2]{#2}\begingroup\raggedright\begin{thebibliography}{10}

\bibitem{Ade:2015xua}
{\bf Planck} Collaboration, P.~A.~R. Ade et~al., {\it {Planck 2015 results.
  XIII. Cosmological parameters}},  {\em Astron. Astrophys.} {\bf 594} (2016)
  A13, [\href{http://arxiv.org/abs/1502.01589}{{\tt arXiv:1502.01589}}].

\bibitem{Costanzi:2018xql}
M.~Costanzi et~al., {\it {Dark Energy Survey Year 1 Results: Methods for
  Cluster Cosmology and Application to the SDSS}},
  \href{http://arxiv.org/abs/1810.09456}{{\tt arXiv:1810.09456}}.

\bibitem{2012PhRvD..86j3518P}
D.~{Parkinson}, S.~{Riemer-S{\o}rensen}, C.~{Blake}, G.~B. {Poole}, T.~M.
  {Davis}, S.~{Brough}, M.~{Colless}, C.~{Contreras}, W.~{Couch}, S.~{Croom},
  D.~{Croton}, M.~J. {Drinkwater}, K.~{Forster}, D.~{Gilbank}, M.~{Gladders},
  K.~{Glazebrook}, B.~{Jelliffe}, R.~J. {Jurek}, I.-h. {Li}, B.~{Madore}, D.~C.
  {Martin}, K.~{Pimbblet}, M.~{Pracy}, R.~{Sharp}, E.~{Wisnioski}, D.~{Woods},
  T.~K. {Wyder}, and H.~K.~C. {Yee}, {\it {The WiggleZ Dark Energy Survey:
  Final data release and cosmological results}},  {\em Phys. Rev. D} {\bf 86}
  (Nov., 2012) 103518, [\href{http://arxiv.org/abs/1210.2130}{{\tt
  arXiv:1210.2130}}].

\bibitem{Tegmark:2006az}
{\bf SDSS} Collaboration, M.~Tegmark et~al., {\it {Cosmological Constraints
  from the SDSS Luminous Red Galaxies}},  {\em Phys. Rev.} {\bf D74} (2006)
  123507, [\href{http://arxiv.org/abs/astro-ph/0608632}{{\tt
  astro-ph/0608632}}].

\bibitem{Tegmark:2003ud}
{\bf SDSS} Collaboration, M.~Tegmark et~al., {\it {Cosmological parameters from
  SDSS and WMAP}},  {\em Phys. Rev.} {\bf D69} (2004) 103501,
  [\href{http://arxiv.org/abs/astro-ph/0310723}{{\tt astro-ph/0310723}}].

\bibitem{Clowe:2006eq}
D.~Clowe, M.~Bradac, A.~H. Gonzalez, M.~Markevitch, S.~W. Randall, C.~Jones,
  and D.~Zaritsky, {\it {A direct empirical proof of the existence of dark
  matter}},  {\em Astrophys. J.} {\bf 648} (2006) L109--L113,
  [\href{http://arxiv.org/abs/astro-ph/0608407}{{\tt astro-ph/0608407}}].

\bibitem{Begeman:1991iy}
K.~G. Begeman, A.~H. Broeils, and R.~H. Sanders, {\it {Extended rotation curves
  of spiral galaxies: Dark haloes and modified dynamics}},  {\em Mon. Not. Roy.
  Astron. Soc.} {\bf 249} (1991) 523.

\bibitem{Heeba:2018wtf}
S.~Heeba, F.~Kahlhoefer, and P.~Stöcker, {\it {Freeze-in production of
  decaying dark matter in five steps}},
  \href{http://arxiv.org/abs/1809.04849}{{\tt arXiv:1809.04849}}.

\bibitem{Steigman:2012nb}
G.~Steigman, B.~Dasgupta, and J.~F. Beacom, {\it {Precise Relic WIMP Abundance
  and its Impact on Searches for Dark Matter Annihilation}},  {\em Phys. Rev.}
  {\bf D86} (2012) 023506, [\href{http://arxiv.org/abs/1204.3622}{{\tt
  arXiv:1204.3622}}].

\bibitem{Carlson:1992fn}
E.~D. Carlson, M.~E. Machacek, and L.~J. Hall, {\it {Self-interacting dark
  matter}},  {\em Astrophys. J.} {\bf 398} (1992) 43--52.

\bibitem{Okun:1980kw}
L.~B. Okun, {\it {THETONS}},  {\em JETP Lett.} {\bf 31} (1980) 144--147. [Pisma
  Zh. Eksp. Teor. Fiz.31,156(1979)].

\bibitem{Faraggi:2000pv}
A.~E. Faraggi and M.~Pospelov, {\it {Selfinteracting dark matter from the
  hidden heterotic string sector}},  {\em Astropart. Phys.} {\bf 16} (2002)
  451--461, [\href{http://arxiv.org/abs/hep-ph/0008223}{{\tt hep-ph/0008223}}].

\bibitem{Juknevich:2009ji}
J.~E. Juknevich, D.~Melnikov, and M.~J. Strassler, {\it {A Pure-Glue Hidden
  Valley I. States and Decays}},  {\em JHEP} {\bf 07} (2009) 055,
  [\href{http://arxiv.org/abs/0903.0883}{{\tt arXiv:0903.0883}}].

\bibitem{Acharya:2017szw}
B.~S. Acharya, M.~Fairbairn, and E.~Hardy, {\it {Glueball dark matter in
  non-standard cosmologies}},  {\em JHEP} {\bf 07} (2017) 100,
  [\href{http://arxiv.org/abs/1704.01804}{{\tt arXiv:1704.01804}}].

\bibitem{NUSSINOV198555}
S.~Nussinov, {\it Technocosmology — could a technibaryon excess provide a
  “natural” missing mass candidate?},  {\em Physics Letters B} {\bf 165}
  (1985), no.~1 55 -- 58.

\bibitem{BARR1990387}
S.~Barr, R.~S. Chivukula, and E.~Farhi, {\it Electroweak fermion number
  violation and the production of stable particles in the early universe},
  {\em Physics Letters B} {\bf 241} (1990), no.~3 387 -- 391.

\bibitem{Barr:1991qn}
S.~M. Barr, {\it {Baryogenesis, sphalerons and the cogeneration of dark
  matter}},  {\em Phys. Rev.} {\bf D44} (1991) 3062--3066.

\bibitem{PhysRevLett.68.741}
D.~B. Kaplan, {\it Single explanation for both baryon and dark matter
  densities},  {\em Phys. Rev. Lett.} {\bf 68} (Feb, 1992) 741--743.

\bibitem{Kaplan:2009ag}
D.~E. Kaplan, M.~A. Luty, and K.~M. Zurek, {\it {Asymmetric Dark Matter}},
  {\em Phys. Rev.} {\bf D79} (2009) 115016,
  [\href{http://arxiv.org/abs/0901.4117}{{\tt arXiv:0901.4117}}].

\bibitem{Kribs:2009fy}
G.~D. Kribs, T.~S. Roy, J.~Terning, and K.~M. Zurek, {\it {Quirky Composite
  Dark Matter}},  {\em Phys. Rev.} {\bf D81} (2010) 095001,
  [\href{http://arxiv.org/abs/0909.2034}{{\tt arXiv:0909.2034}}].

\bibitem{Blennow:2010qp}
M.~Blennow, B.~Dasgupta, E.~Fernandez-Martinez, and N.~Rius, {\it {Aidnogenesis
  via Leptogenesis and Dark Sphalerons}},  {\em JHEP} {\bf 03} (2011) 014,
  [\href{http://arxiv.org/abs/1009.3159}{{\tt arXiv:1009.3159}}].

\bibitem{2011PhLB..701..296M}
G.~{Mangano} and P.~D. {Serpico}, {\it {A robust upper limit on N$_{}$ from
  BBN, circa 2011}},  {\em Physics Letters B} {\bf 701} (July, 2011) 296--299,
  [\href{http://arxiv.org/abs/1103.1261}{{\tt arXiv:1103.1261}}].

\bibitem{Percival:2006gt}
W.~J. Percival et~al., {\it {The shape of the SDSS DR5 galaxy power spectrum}},
   {\em Astrophys. J.} {\bf 657} (2007) 645--663,
  [\href{http://arxiv.org/abs/astro-ph/0608636}{{\tt astro-ph/0608636}}].

\bibitem{sz1970}
R.~A. {Sunyaev} and Y.~B. {Zeldovich}, {\it {Small-Scale Fluctuations of Relic
  Radiation}},  {\em \apss} {\bf 7} (Apr., 1970) 3--19.

\bibitem{eh1998}
D.~J. {Eisenstein} and W.~{Hu}, {\it {Baryonic Features in the Matter Transfer
  Function}},  {\em \apj} {\bf 496} (Mar., 1998) 605--614,
  [\href{http://arxiv.org/abs/astro-ph/9709112}{{\tt astro-ph/9709112}}].

\bibitem{Kamada:2016qjo}
A.~Kamada, K.~Kohri, T.~Takahashi, and N.~Yoshida, {\it {Effects of
  electrically charged dark matter on cosmic microwave background
  anisotropies}},  {\em Phys. Rev.} {\bf D95} (2017), no.~2 023502,
  [\href{http://arxiv.org/abs/1604.07926}{{\tt arXiv:1604.07926}}].

\bibitem{Kamada:2017oxi}
A.~Kamada and T.~Takahashi, {\it {Dark matter kinetic decoupling with a light
  particle}},  {\em JCAP} {\bf 1801} (2018), no.~01 047,
  [\href{http://arxiv.org/abs/1703.02338}{{\tt arXiv:1703.02338}}].

\bibitem{Sarkar:2017vls}
A.~Sarkar, S.~K. Sethi, and S.~Das, {\it {The effects of the small-scale
  behaviour of dark matter power spectrum on CMB spectral distortion}},  {\em
  JCAP} {\bf 1707} (2017), no.~07 012,
  [\href{http://arxiv.org/abs/1701.07273}{{\tt arXiv:1701.07273}}].

\bibitem{Boehm:2000gq}
C.~Boehm, P.~Fayet, and R.~Schaeffer, {\it {Constraining dark matter candidates
  from structure formation}},  {\em Phys. Lett.} {\bf B518} (2001) 8--14,
  [\href{http://arxiv.org/abs/astro-ph/0012504}{{\tt astro-ph/0012504}}].

\bibitem{Boehm:2001hm}
C.~Boehm, A.~Riazuelo, S.~H. Hansen, and R.~Schaeffer, {\it {Interacting dark
  matter disguised as warm dark matter}},  {\em Phys. Rev.} {\bf D66} (2002)
  083505, [\href{http://arxiv.org/abs/astro-ph/0112522}{{\tt
  astro-ph/0112522}}].

\bibitem{Chen:2002yh}
X.-l. Chen, S.~Hannestad, and R.~J. Scherrer, {\it {Cosmic microwave background
  and large scale structure limits on the interaction between dark matter and
  baryons}},  {\em Phys. Rev.} {\bf D65} (2002) 123515,
  [\href{http://arxiv.org/abs/astro-ph/0202496}{{\tt astro-ph/0202496}}].

\bibitem{Sigurdson:2003vy}
K.~Sigurdson and M.~Kamionkowski, {\it {Charged - particle decay and
  suppression of small - scale power}},  {\em Phys. Rev. Lett.} {\bf 92} (2004)
  171302, [\href{http://arxiv.org/abs/astro-ph/0311486}{{\tt
  astro-ph/0311486}}].

\bibitem{Nusser:2004qu}
A.~Nusser, S.~S. Gubser, and P.~J.~E. Peebles, {\it {Structure formation with a
  long-range scalar dark matter interaction}},  {\em Phys. Rev.} {\bf D71}
  (2005) 083505, [\href{http://arxiv.org/abs/astro-ph/0412586}{{\tt
  astro-ph/0412586}}].

\bibitem{Das:2006ht}
S.~Das and N.~Weiner, {\it {Late Forming Dark Matter in Theories of Neutrino
  Dark Energy}},  {\em Phys. Rev.} {\bf D84} (2011) 123511,
  [\href{http://arxiv.org/abs/astro-ph/0611353}{{\tt astro-ph/0611353}}].

\bibitem{Aarssen:2012fx}
L.~G. van~den Aarssen, T.~Bringmann, and C.~Pfrommer, {\it {Is dark matter with
  long-range interactions a solution to all small-scale problems of
  $\Lambda$CDM cosmology?}},  {\em Phys. Rev. Lett.} {\bf 109} (2012) 231301,
  [\href{http://arxiv.org/abs/1205.5809}{{\tt arXiv:1205.5809}}].

\bibitem{Wilkinson:2013kia}
R.~J. Wilkinson, J.~Lesgourgues, and C.~Boehm, {\it {Using the CMB angular
  power spectrum to study Dark Matter-photon interactions}},  {\em JCAP} {\bf
  1404} (2014) 026, [\href{http://arxiv.org/abs/1309.7588}{{\tt
  arXiv:1309.7588}}].

\bibitem{Wilkinson:2014ksa}
R.~J. Wilkinson, C.~Boehm, and J.~Lesgourgues, {\it {Constraining Dark
  Matter-Neutrino Interactions using the CMB and Large-Scale Structure}},  {\em
  JCAP} {\bf 1405} (2014) 011, [\href{http://arxiv.org/abs/1401.7597}{{\tt
  arXiv:1401.7597}}].

\bibitem{Chu:2014lja}
X.~Chu and B.~Dasgupta, {\it {Dark Radiation Alleviates Problems with Dark
  Matter Halos}},  {\em Phys. Rev. Lett.} {\bf 113} (2014), no.~16 161301,
  [\href{http://arxiv.org/abs/1404.6127}{{\tt arXiv:1404.6127}}].

\bibitem{Buckley:2014hja}
M.~R. Buckley, J.~Zavala, F.-Y. Cyr-Racine, K.~Sigurdson, and M.~Vogelsberger,
  {\it {Scattering, Damping, and Acoustic Oscillations: Simulating the
  Structure of Dark Matter Halos with Relativistic Force Carriers}},  {\em
  Phys. Rev.} {\bf D90} (2014), no.~4 043524,
  [\href{http://arxiv.org/abs/1405.2075}{{\tt arXiv:1405.2075}}].

\bibitem{Archidiacono:2015oma}
M.~Archidiacono, S.~Hannestad, R.~S. Hansen, and T.~Tram, {\it {Sterile
  neutrinos with pseudoscalar self-interactions and cosmology}},  {\em Phys.
  Rev.} {\bf D93} (2016), no.~4 045004,
  [\href{http://arxiv.org/abs/1508.02504}{{\tt arXiv:1508.02504}}].

\bibitem{Cyr-Racine:2015ihg}
F.-Y. Cyr-Racine, K.~Sigurdson, J.~Zavala, T.~Bringmann, M.~Vogelsberger, and
  C.~Pfrommer, {\it {ETHOS—an effective theory of structure formation: From
  dark particle physics to the matter distribution of the Universe}},  {\em
  Phys. Rev.} {\bf D93} (2016), no.~12 123527,
  [\href{http://arxiv.org/abs/1512.05344}{{\tt arXiv:1512.05344}}].

\bibitem{Chacko:2016kgg}
Z.~Chacko, Y.~Cui, S.~Hong, T.~Okui, and Y.~Tsai, {\it {Partially Acoustic Dark
  Matter, Interacting Dark Radiation, and Large Scale Structure}},  {\em JHEP}
  {\bf 12} (2016) 108, [\href{http://arxiv.org/abs/1609.03569}{{\tt
  arXiv:1609.03569}}].

\bibitem{Dror:2017gjq}
J.~A. Dror, E.~Kuflik, B.~Melcher, and S.~Watson, {\it {Concentrated dark
  matter: Enhanced small-scale structure from codecaying dark matter}},  {\em
  Phys. Rev.} {\bf D97} (2018), no.~6 063524,
  [\href{http://arxiv.org/abs/1711.04773}{{\tt arXiv:1711.04773}}].

\bibitem{Buen-Abad:2018mas}
M.~A. Buen-Abad, R.~Emami, and M.~Schmaltz, {\it {Cannibal Dark Matter and
  Large Scale Structure}},  \href{http://arxiv.org/abs/1803.08062}{{\tt
  arXiv:1803.08062}}.

\bibitem{Hu:1998kj}
W.~Hu, {\it {Structure formation with generalized dark matter}},  {\em
  Astrophys. J.} {\bf 506} (1998) 485--494,
  [\href{http://arxiv.org/abs/astro-ph/9801234}{{\tt astro-ph/9801234}}].

\bibitem{Dodelson:2003ft}
S.~Dodelson, {\em {Modern Cosmology}}.
\newblock Academic Press, Amsterdam, 2003.

\bibitem{Ma:1995ey}
C.-P. Ma and E.~Bertschinger, {\it {Cosmological perturbation theory in the
  synchronous and conformal Newtonian gauges}},  {\em Astrophys. J.} {\bf 455}
  (1995) 7--25, [\href{http://arxiv.org/abs/astro-ph/9506072}{{\tt
  astro-ph/9506072}}].

\bibitem{Lewis:1999bs}
A.~Lewis, A.~Challinor, and A.~Lasenby, {\it {Efficient computation of CMB
  anisotropies in closed FRW models}},  {\em Astrophys. J.} {\bf 538} (2000)
  473--476, [\href{http://arxiv.org/abs/astro-ph/9911177}{{\tt
  astro-ph/9911177}}].

\bibitem{2011JCAP...07..034B}
D.~{Blas}, J.~{Lesgourgues}, and T.~{Tram}, {\it {The Cosmic Linear Anisotropy
  Solving System (CLASS). Part II: Approximation schemes}},  {\em JCAP} {\bf 7}
  (July, 2011) 034, [\href{http://arxiv.org/abs/1104.2933}{{\tt
  arXiv:1104.2933}}].

\bibitem{df1992}
A.~D. {Dolgov} and M.~{Fukugita}, {\it {Nonequilibrium effect of the neutrino
  distribution on primordial helium synthesis}},  {\em \prd} {\bf 46} (Dec.,
  1992) 5378--5382.

\bibitem{hm1995}
S.~{Hannestad} and J.~{Madsen}, {\it {Neutrino decoupling in the early
  Universe}},  {\em \prd} {\bf 52} (Aug., 1995) 1764--1769,
  [\href{http://arxiv.org/abs/astro-ph/9506015}{{\tt astro-ph/9506015}}].

\bibitem{dhs1997}
A.~D. {Dolgov}, S.~H. {Hansen}, and D.~V. {Semikoz}, {\it {Non-equilibrium
  corrections to the spectra of massless neutrinos in the early universe}},
  {\em Nuclear Physics B} {\bf 503} (Feb., 1997) 426--444,
  [\href{http://arxiv.org/abs/hep-ph/9703315}{{\tt hep-ph/9703315}}].

\bibitem{fks1997}
N.~{Fornengo}, C.~W. {Kim}, and J.~{Song}, {\it {Finite temperature effects on
  the neutrino decoupling in the early Universe}},  {\em \prd} {\bf 56} (Oct.,
  1997) 5123--5134, [\href{http://arxiv.org/abs/hep-ph/9702324}{{\tt
  hep-ph/9702324}}].

\bibitem{mmpp2002}
G.~{Mangano}, G.~{Miele}, S.~{Pastor}, and M.~{Peloso}, {\it {A precision
  calculation of the effective number of cosmological neutrinos}},  {\em
  Physics Letters B} {\bf 534} (May, 2002) 8--16,
  [\href{http://arxiv.org/abs/astro-ph/0111408}{{\tt astro-ph/0111408}}].

\bibitem{Cooke:2013cba}
R.~Cooke, M.~Pettini, R.~A. Jorgenson, M.~T. Murphy, and C.~C. Steidel, {\it
  {Precision measures of the primordial abundance of deuterium}},  {\em
  Astrophys. J.} {\bf 781} (2014), no.~1 31,
  [\href{http://arxiv.org/abs/1308.3240}{{\tt arXiv:1308.3240}}].

\bibitem{Abazajian:2016yjj}
{\bf CMB-S4} Collaboration, K.~N. Abazajian et~al., {\it {CMB-S4 Science Book,
  First Edition}},  \href{http://arxiv.org/abs/1610.02743}{{\tt
  arXiv:1610.02743}}.

\end{thebibliography}\endgroup
\end{document}